\documentclass[letterpaper,showpacs,aps,prd]{revtex4-2}

\usepackage[utf8]{inputenc}
\usepackage[english]{babel}

\usepackage{latexsym}
\usepackage{amssymb}
\usepackage{amsmath}
\usepackage{bm} 
\usepackage{tabularx}
\usepackage{graphicx}
\usepackage{epsfig}
\usepackage{caption}
\usepackage[justification=raggedright]{subcaption}
\usepackage[colorlinks=true, linkcolor=blue, citecolor=blue, urlcolor=blue]{hyperref}
\usepackage{comment}

\newcommand{\mathper}[1]{\text{\slshape #1}}

\usepackage{color}

\begin{document}

\captionsetup[figure]{labelfont=bf,singlelinecheck=off,justification=raggedright}


\title{Electrically Charged Proca Stars}

\author{Yahir Mio}
\email{jorge.mio@correo.nucleares.unam.mx}

\author{Miguel Alcubierre}
\email{malcubi@nucleares.unam.mx}

\affiliation{Instituto de Ciencias Nucleares, Universidad Nacional
Aut\'onoma de M\'exico, Circuito exterior C.U., A.P. 70-543, Ciudad de M\'exico 04510, M\'exico.}


\date{\today}




\begin{abstract}
We consider self-gravitating stationary configurations of a charged massive complex Proca field, also known as ``charged Proca stars", in the particular case of spherical symmetry. We first present a general 3+1 decomposition of the Einstein--Maxwell--Proca system, starting from the action and field equations.  We then restrict our system to the case of spherical symmetry and, after imposing a harmonic time dependence ansatz for the Proca field, we construct families of charged Proca stars for different values of the charge parameter $q$, and different values of the central Proca scalar potential $\varphi$. In a similar way to the case of scalar boson stars, one can define a critical charge $q=q_c$ that corresponds to the value for which the Coulomb repulsion of the charged Proca field exactly cancels their newtonian gravitational attraction. Just as in the case of boson stars studied in~\cite{Pugliese_2013,Lopez:2023phk}, we find that supercritical solutions can exist for a limited range of charges above the critical value $q>q_c$.  We also consider the binding energy $E_B$ for the different families of solutions, and find that gravitationally bound solutions such that $E_B<0$ can only exist for subcritical charges such that $q<q_c$, indicating that our supercritical solutions are probably dynamically unstable against perturbations.
\end{abstract}


\maketitle


\section{Introduction}

Proca stars are a particular type of self-gravitating exotic compact object (ECO) formed from a massive complex vector field with spin $1$, the so-called Proca field, minimally coupled to gravity, and as such they are solutions of the Einstein--Proca system of equations~\cite{Brito:2015pxa,SanchisGual:2017,SanchisGual:2019,SanchisGual:2022,DiGiovanni:2018bvo}. The Proca star model was introduced as a natural extension to the vector case of the usual scalar boson stars, initially proposed by Kaup and Ruffini in~\cite{Kaup68,Ruffini69} (for a detailed review of these types of objects see~\cite{Liebling:2012fv,Herdeiro:2020jzx}).  More recently, interesting variants of the original Proca star model have also been studied, for example adding a self-interaction potential~\cite{Aoki:2022mdn}, considering more than one Proca field~\cite{Zhang:2023rwc,Lazarte:2024jyr,Lazarte:2025wlw}, or even including other matter fields~\cite{Ma:2023vfa,Pombo:2023sih,Jockel:2023rrm,Ma:2023bhb,Herdeiro:2023a,Herdeiro:2024b}.

Research on Proca stars (and boson stars) has some astrophysical motivations: They have been proposed as candidates for dark matter~\cite{Hernandez:2023tig}, and also as possible black hole mimickers~\cite{Rosa_2022,Sengo:2022jif,Sengo_2024}. Although to date all the detected gravitational wave signals have been explained by the collision of known objects such as black holes or neutron stars, it is worth mentioning the fact that the gravitational wave event GW190521~\cite{Abbott:2020}, usually interpreted as a merger of two black holes, was later reanalyzed in~\cite{CalderonBustillo:2020fyi} where they find that the characteristics of this signal are also consistent with the merger of two Proca stars. The study of Proca stars and other ECO's is also important from a purely theoretical perspective: in addition to being interesting solutions in general relativity, they can serve as toy models to understand the interaction between gravity and different field theories~\cite{Torres97}.

In this work, we study the case of electrically charged Proca stars (CPS) in a static and spherically symmetric spacetime. These solutions describe ECO's in which the Proca and Maxwell fields are coupled to each other, and minimally coupled to the surrounding spacetime, resulting in the Einstein--Maxwell--Proca (EMP) system of equations. Several studies have already considered solutions for electrically charged scalar boson stars~\cite{Lee:1989,Jetzer89,Pugliese_2013,Lopez:2023phk}, where families of such solutions have been studied for different values of the charge parameter $q$, up to and even slightly beyond the critical value $q_c=m$, where $m$ is the mass parameter of the scalar field.  However, as far as we know, the only study of charged Proca stars so far has been that of Landea and García in~\cite{SalazarLandea:2016bys}, although they only reported solutions for a very specific value of $q$.

In this work, we construct families of charged Proca stars with different values of the charge parameter $q$. In order to solve the EMP system of equations we have first consider the 3+1 decomposition of the Maxwell and Proca fields, which together with the 3+1 decomposition of the gravitational field given for example in~\cite{Alcubierre08a}, allows us to obtain a complete formulation of the EMP system in the 3+1 formalism.  We then restrict our system to the case of spherical symmetry, and after imposing the Proca star ansatz solve the resulting eigenvalue problem numerically. It is interesting to note that, just as it was shown for the case of scalar boson stars in~\cite{Lopez:2023phk}, we are also able to find some supercritical solutions such that $q>q_c$.

This paper is organized as follows. In Section II we present the 3+1 decomposition of the EMP system, starting from the action and field equations. In Section III we particularize the system of equations to the case of spherically symmetry. In Section IV we derive the radial equations corresponding to electrostatic configurations for charged Proca stars. In Section V we present the boundary conditions for the system, as well as the necessary rescaling and gauge transformations needed to obtain the physical frequencies. In Section VI we describe how we can calculate the total mass and charge of our configurations. Section VII presents the numerical results for both the different families of stationary solutions, as well as the radial profile of the different variables for some particular solutions. We conclude in Section VIII.


\section{The Einstein-Maxwell-Proca system in 3+1 formalism}


\subsection{Action and field equations}

We consider a complex Proca field coupled to a real Maxwell field, both minimally coupled to gravity. The system is described by the following action (throughout this paper we use units such that $G = c =\hbar = 1$, and the signature of the spacetime metric is taken to be $(-,+,+,+)$):
\begin{equation} \label{accion-general}
S = \int \frac{1}{16\pi} \left[ R - \bar{\mathcal{W}}^{\mu\nu} \mathcal{W}_{\mu\nu}
- 2m^2\bar{X}^{\mu}X_\mu - F^{\mu\nu} F_{\mu\nu} \right]
\sqrt{-g} \: d^4 x \; ,
\end{equation}
where $g$ is the determinant of the spacetime metric, $R$ is the curvature scalar of our spacetime, $X_\mu$ the Proca potential 1-form (with the overbar denoting the complex conjugate), $\mathcal{W}_{\mu \nu}$ the Proca field tensor, $F_{\mu \nu}$ the Maxwell (Faraday) field tensor, and $m$ the mass parameter of the Proca field. The Proca field tensor $\mathcal{W}_{\mu \nu}$ is defined in terms of the Proca potential as $\mathcal{W}_{\mu\nu} := \mathcal{D}_\mu X_\nu - \mathcal{D}_\nu X_\mu$, where $\mathcal{D}_\mu := \nabla_\mu - iqA_\mu$ denotes the gauge invariant derivative which includes the coupling to the Maxwell potential 1-form $A_\mu$, with $q$ the corresponding electric charge parameter of the Proca field. On the other hand, the Maxwell field $F_{\mu \nu}$ is defined in terms of the potential 1-form $A_\mu$ in the usual way, $F_{\mu\nu} := \nabla_\mu A_\nu - \nabla_\nu A_\mu$.

Variation of the action with respect to the metric $g_{\mu \nu}$, the Maxwell potential $A_\mu$, and Proca potential $X_\mu$, yields the following field equations:
\begin{align}
R_{\mu\nu} - \frac{1}{2} \: g_{\mu\nu} R &= 8 \pi T_{\mu\nu}\; ,
\label{eq-Einstein} \\
\nabla_\alpha F^{\alpha\mu} &= - 4 \pi j^\mu \; ,
\label{eq-Maxwell} \\
\mathcal{D}_\alpha \mathcal{W}^{\alpha\beta} &= m^2 X^\beta \; ,
\label{eq-Proca}
\end{align}
with $R_{\mu \nu}$ the Ricci tensor of the spacetime, and where $T_{\mu\nu}$ and $j^{\mu}$ are the stress--energy tensor and the electric current density which take the form:
\begin{align}
T_{\mu\nu} &= \frac{1}{4\pi} \left[ -\mathcal{W}_{\lambda(\mu} \bar{\mathcal{W}}_{\nu)}^{\phantom{\nu}\lambda} - \frac{g_{\mu\nu}}{4}  \: \mathcal{W}_{\alpha\beta} \bar{\mathcal{W}}^{\alpha\beta}
+ m^2 \left( X_{(\mu} \bar{X}_{\nu)} - \frac{g_{\mu\nu}}{2} \: X_\lambda \bar{X}^\lambda \right) \right] \nonumber \\
&+ \frac{1}{4\pi} \left[ - F_{\lambda\mu} F_{\nu}^{\phantom{\nu} \lambda}
- \frac{g_{\mu\nu}}{4} \: F_{\alpha\beta} F^{\alpha\beta} \right] \; , \\
j^\mu &= \frac{iq}{8\pi} \left(\bar{\mathcal{W}}^{\mu\beta} X_\beta - \mathcal{W}^{\mu\beta} \bar{X}_\beta\right) \; .
\end{align}
In the stress-energy tensor $T_{\mu \nu}$ above the terms inside the first bracket come from the Lagrangian terms associated with the Proca field, while those in the second bracket come from the terms related to the Maxwell field. In addition, $j^{\mu}$ can be shown to be a conserved current associated with the invariance of the Lagrangian under the transformation: $X_\mu \rightarrow e^{-i q \theta} X_\mu$, $A_\mu \rightarrow A_\mu -\partial_\mu \theta$, with $\theta=\theta(x^\alpha)$ an arbitrary function of space and time. One can also verify that both the stress--energy tensor and the current density satisfy the conservation equations: $\nabla_\mu T^{\mu\nu} = 0$, $\nabla_\mu j^\mu = 0$.

We need to write some additional equations. First, using the gauge freedom of the Maxwell potential $A_\mu$, from now on we will impose the Lorenz condition:
\begin{equation} 
\nabla_\alpha A^\alpha = 0 \; , 
\end{equation}
On the other hand, from the Proca field equation~\eqref{eq-Proca} one can show that $X_\mu$ must satisfy the following condition:
\begin{equation}
\mathcal{D}_\alpha X^\alpha = \frac{iq}{2 m^2} \:
\mathcal{W}^{\alpha\beta} F_{\alpha\beta} \; .
\end{equation}
Notice that for $q=0$ this is essentially a Lorenz condition for the Proca potential.  However, in this case this is not a gauge choice, but it is instead a dynamical requirement.

Additionally, the Proca and Maxwell fields each satisfy their own version of the Bianchi identities, which can be written as:
\begin{equation}
\nabla_\alpha F^{*\alpha\beta} = 0 \; , \qquad
\mathcal{D}_\alpha \mathcal{W}^{*\alpha\beta} = - iq X_\alpha F^{*\alpha\beta} \; ,
\end{equation}
where we have introduced the dual tensor $Z^*_{\alpha\beta} := - \frac{1}{2} \: \tilde{\varepsilon}_{\alpha\beta \mu\nu} Z^{\mu\nu},$ for $Z=(F,\mathcal{W})$, with $\tilde{\varepsilon}_{\alpha\beta\mu\nu}$ the Levi--Civita tensor which is in turn defined in terms of the Levi--Civita symbol $\varepsilon_{\alpha\beta\mu\nu}$ as $\tilde{\varepsilon}_{\alpha\beta\mu\nu} := \sqrt{-g} \: \varepsilon_{\alpha\beta\mu\nu}$ (we use a convention such that $\epsilon_{0123} = 1 = - \epsilon^{0123}$).


\subsection{The Proca equations in the 3+1 formalism}

In order to develop a formalism that will later allow us to study time evolutions, we will work within the framework of the 3+1 formalism of general relativity, where the spacetime ($\mathcal{M}$, $g_{\mu \nu}$) can be foliated by Cauchy hypersurfaces $\Sigma_{t}$ parametrized by a global time function $t$, resulting in the standard Arnowitt--Deser--Misner (ADM) equations (see for example~\cite{Alcubierre08a}). The 3+1 split of the metric takes the following form:
\begin{equation}
ds^2 = \left( - \alpha^2 dt^2 + \beta_i \beta^i \right) dt^2 + 2\beta_i dtdx^i
+ \gamma_{ij} dx^i dx^j \;, \label{GeneralLineElement}
\end{equation}
where $\gamma_{ij}$ is the induced three-dimensional metric defined within the spatial hypersurfaces, $\alpha$ is the lapse function, and $\beta^i$ is the shift vector. From here on, we will adopt the convention that Greek indices take values from 0 to 3, while Latin indices only take values from 1 to 3.

In terms of the lapse and shift, the unit normal vector $n^{\mu}$ to the spatial hypersurfaces has components given by
\begin{equation}
n^\mu = \left( 1/ \alpha, -\beta^{i}/\alpha\right) \; , \qquad
n_\mu = \left(  -\alpha, 0,0,0 \right) \; .
\label{VectorEuler}
\end{equation}

Starting from this, we can decompose any tensor field into its components normal and parallel to the spatial hypersurfaces by projecting it with the normal unit vector $n^\mu$ and the projector operator $\gamma^\nu_\mu := \delta^\nu_\mu - n^\nu n_\mu$, respectively.  Notice that thus defined $\gamma_{ij}$ is precisely the induced metric on the spatial hypersurfaces mentioned above. From the Maxwell and Proca potential 1-forms we then define the ``scalar" and ``3-vector" potentials as:
\begin{alignat}{2}
\varphi &:= - n^\nu X_\nu \; , & \qquad
x_\mu &:= \gamma_\mu^\nu X_\nu \; , \\
\phi &:= - n^\nu A_\nu \; , &
a_\mu &:= \gamma_\mu^\nu A_\nu \; .
\end{alignat}
In particular, these definitions imply that $n^\mu x_\mu = n^\mu a_\mu = 0$, that is $a_\mu$ and $x_\mu$ are purely spatial vectors, so from now on we will only consider their spatial components.  In the same way, from the Maxwell and Proca field tensors we can define their ``electric" and ``magnetic" parts as:
\begin{alignat}{2}
\mathcal{E}_\mu &:= n^\nu \mathcal{W}_{\mu\nu} \; , & \qquad
\mathcal{B}_\mu &:= n^\nu \mathcal{W}^*_{\mu\nu} \; , \\
E_\mu &:= n^\nu F_{\mu\nu} \; , &
B_\mu &:= n^\nu F^*_{\mu\nu} \; .
\end{alignat}
Again, one can easily see that $n^\mu E_\mu = n^\mu \mathcal{E}_\mu = 0$,
and $n^\mu B_\mu = n^\mu \mathcal{B}_\mu = 0$. From the above definitions one can immediately recover the original 1-forms and field tensors as:
\begin{align}
X_\mu &= x_\mu + n_\mu \varphi \; , \\
A_\mu &= a_\mu + n_\mu\phi \; ,
\end{align}
and:
\begin{align}
\mathcal{W}_{\mu\nu} &= n_\mu \mathcal{E}_\nu - n_\nu \mathcal{E}_\mu
+ \tilde{\varepsilon}_{\mu\nu\lambda} \mathcal{B}^\lambda \; , \\
\mathcal{W}^*_{\mu\nu} &= n_\mu \mathcal{B}_\nu - n_\nu \mathcal{B}_\mu
- \tilde{\varepsilon}_{\mu\nu\lambda} \mathcal{E}^\lambda \; , \\
F_{\mu\nu} &= n_\mu E_\nu - n_\nu E_\mu + \tilde{\varepsilon}_{\mu\nu\lambda} B^\lambda \; , \\
F^*_{\mu\nu} &= n_\mu B_\nu - n_\nu B_\mu - \tilde{\varepsilon}_{\mu\nu\lambda} E^\lambda \; ,
\end{align}
where the three-dimensional Levi--Civita tensor is now defined as $\tilde{\varepsilon}_{\mu\nu\lambda} := n^\alpha \tilde{\varepsilon}_{\alpha \mu \nu \lambda}$.

The decomposition of the matter terms is achieved with projections of the stress--energy tensor $T_{\mu \nu}$. The energy density is defined as the double normal projection, $\rho := n^\mu n^\nu T_{\mu\nu}$, the momentum density as a mixed projection, $J^i := - \gamma^{i\mu} n^\nu T_{\mu\nu}$, and the spatial stress tensor as the double projection onto the hypersurface, $S_{ij} := \gamma^\mu_i \gamma^\nu_j T_{\mu\nu}$. In terms of the potentials and fields these quantities become:
\begin{align}
\rho &= \frac{1}{8\pi} \left[ \mathcal{E}^2 + \mathcal{B}^2 + m^2 \left( \varphi^2 + x^2 \right)  \right] +\frac{1}{8\pi} \left [E^2 + B^2 \right] \; , \\
J^i &=  \frac{1}{8\pi} \left[  \left(\bar{\mathcal{E}} \times \mathcal{B}\right)^i + m^2 x^i\bar{\varphi} + c.c.\right] + \frac{1}{4\pi} \left(E \times B\right)^i \; , \\
S_{ij} &= \frac{1}{8\pi}	 \left\{ \gamma_{ij} \left(\mathcal{E}^2 + \mathcal{B}^2\right) - \left(\mathcal{B}_i \bar{\mathcal{B}}_j + \mathcal{E}_i \bar{\mathcal{E}}_j + c.c. \right) + m^2 \left[ \left(x_i \bar{x}_j + c.c. \right) - \gamma_{ij} \left(x^2 - \varphi^2 \right)\right] \right\} \notag \\
&+ \frac{1}{8\pi}	 \left\{\gamma_{ij} \left(E^2 + B^2\right) - 2 \left(B_i B_j + E_i E_j\right) \right\} \; ,
\end{align}
where $c.c.$ means the complex conjugate of the previous expression.

Similarly, we decompose the 4-current density $j^\mu$ into the charge density, $\rho_Q:= n^\mu j_\mu$, and the three-dimensional current density, $j^i_Q = \gamma^i_\mu j^\mu$. We find:
\begin{align}
\rho_Q &= \frac{iq}{8\pi} \left( x_k \bar{\mathcal{E}}^k - c.c \right) \; , \\
j^i_Q &= \frac{iq}{8\pi} \left[ \varphi \bar{\mathcal{E}}^i + \left(x \times \bar{\mathcal{B}}\right)^i - c.c. \right] \; .
\end{align}

Let us now define the three-dimensional gauge invariant derivative as \mbox{$\mathcal{D}^{(3)}_k := \gamma^\mu_k \mathcal{D}_\mu = \nabla^{(3)}_k - iqa_k$}.  From the Maxwell and Proca field equations we can now obtain the following evolution equations for the respective potentials and fields:
\begin{align}
\partial_t \varphi - \pounds_{\vec{\beta}} \varphi &= - \mathcal{D}^{(3)}_i (\alpha x^i) + \alpha\varphi \left(K - iq\phi\right) + \frac{iq\alpha}{m^2} \left( -E_i\mathcal{E}^i + B_i \mathcal{B}^i \right) \; ,
\label{ev-varphi} \\
\partial_tx_i -\pounds_{\vec{\beta}} x_i &= - \mathcal{D}^{(3)}_i (\alpha\varphi) -\alpha\mathcal{E}_i - iq\alpha x_i \phi \; ,
\label{ev-x} \\
\partial_t\mathcal{E}^i - \pounds_{\vec{\beta}} \mathcal{E}^i &= + \left[\mathcal{D}^{(3)} \times (\alpha \mathcal{B})\right]^i + \alpha \left(K - iq\phi\right) \mathcal{E}^i + \alpha m^2x^i \; ,
\label{ev-Ep} \\
\partial_t\mathcal{B}^i - \pounds_{\vec{\beta}} \mathcal{B}^i &= - \left[\mathcal{D}^{(3)} \times (\alpha \mathcal{E}) \right]^i  + \alpha \left(K - iq\phi\right) \mathcal{B}^i + iq\alpha \left[\varphi B^i - \left(x \times E\right)^i\right] \; ,
\label{ev-Bp} \\
\partial_t \phi - \pounds_{\vec{\beta}} \phi &= -\nabla^{(3)}_i (\alpha a^i) +\alpha\phi K \; ,
\label{ev-phi} \\
\partial_t a_i -\pounds_{\vec{\beta}} a_i &= - \nabla^{(3)}_i (\alpha\phi) -\alpha E_i \; ,
\label{ev-a} \\
\partial_t E^i - \pounds_{\vec{\beta}} E^i &= + \left[ \nabla^{(3)}\times(\alpha B) \right]^i
+ \alpha KE^i - 4\pi  \alpha j_Q^i \; ,
\label{ev-Em} \\
\partial_t B^i - \pounds_{\vec{\beta}} B^i &= - \left[ \nabla^{(3)}\times(\alpha E) \right]^i
+ \alpha K B^i \; ,
\label{ev-Bm}
\end{align}
where here $K := {K_m}^m$ is the trace of the extrinsic curvature of the spatial hypersurfaces.
The derivation of these equations is as follows: The evolution equations for the potentials $\varphi$ and $\phi$ emerge from the Lorenz conditions for $X$ and $A$, while the evolution equations for the potentials $x$ and $a$ are obtained from the definitions of $\mathcal{E}$ and $E$, respectively. To derive the remaining evolution equations we project the field equations onto the hypersurface $\Sigma_t$. The equation for $\mathcal{E}$ arises from the projection of the Proca field equation, whereas the equation for $E$ follows from the projection of Maxwell field equation. Similarly, by projecting the dual Proca field equation we obtain the evolution equation for $\mathcal{B}$, while the evolution equation for $B$ is derived from the projection of the dual Maxwell field equation.

Additionally, we have the following set of constraints:
\begin{align}
\nabla^{(3)}_i E^i &= 4 \pi \rho_Q \; , \label{cons-Em} \\
\nabla^{(3)}_i B^i &= 0 \; , \label{cons-Bm} \\
\mathcal{D}^{(3)}_i \mathcal{E}^i &= - m^2 \varphi \; , \label{cons-Ep} \\
D^{(3)}_i \mathcal{B}^i &= - i q x_i B^i \; , \label{cons-Bp}
\end{align}
where the Maxwell and Proca ``electric constraints"~\eqref{cons-Em} and~\eqref{cons-Ep} are obtained by projecting the Maxwell and Proca field equations along the normal direction to the spatial hypersurfaces, while the Maxwell and Proca ``magnetic constraints"~\eqref{cons-Bm} and~\eqref{cons-Bp} are derived by projecting the respective dual field equations.


\section{Reduction to spherical symmetry}

The next step in deriving the equations that model the electrically charged Proca stars is to adapt the equations obtained above for the Einstein-Maxwell-Proca system in the 3+1 formalism to the special case of spherical symmetry. To do this, we consider spherical coordinates ($r,\Theta,\Phi$), and start from a metric of the form:
\begin{equation}
ds^2 = (- \alpha^2 + \beta_r \beta^r) dt^2 + 2 \beta_r dr dt + \mathper{a} \: dr^2
+  r^2 \mathper{b} \: d\Omega^2 \; ,
\end{equation}
where now ($\alpha, \beta^r, \mathper{a}, \mathper{b}$) are functions of $(r,t)$ only, and with $d\Omega^2$ the usual solid angle element \mbox{$d\Omega^2 = d\Theta^2 + \sin^2 \Theta d\Phi^2$}. Notice also that in spherical symmetry the magnetic fields, both Proca and Maxwell, cancel out since \linebreak \mbox{$\mathcal{B}^i = (\nabla^{(3)} \times x)^i - iq (a \times x)^i$} and $B^i = (\nabla^{(3)} \times a)^i$, but both $x$ and $a$ only have radial components.
	
Using the above considerations, we can reduce the expressions for the different quantities for the case of spherical symmetry. In particular, the matter terms become:
\begin{align}
\rho &= \frac{1}{8\pi} \left[\mathcal{E}_r \bar{\mathcal{E}}^r + m^2(\varphi\bar{\varphi} + x_r\bar{x}^r) + E_rE^r\right] \; , \\
J^r &= \frac{m^2}{8\pi} \left(x^r\bar{\varphi} + \bar{x}^r \varphi\right) \; , \\
S_{ij} &= \frac{1}{8\pi} \left\{\gamma_{ij} \left(\mathcal{E}_r \bar{\mathcal{E}}^r + E_r E^r\right)\right. - \left(\mathcal{E}_i \bar{\mathcal{E}}_j + \bar{\mathcal{E}}_i \mathcal{E}_j  + 2 E_i E_j\right) \notag \\
& \left. + m^2 \left[ \left(x_i \bar{x}_j + \bar{x}_i x_j \right) - \gamma_{ij} \left(x_r \bar{x}^r - \varphi \bar{\varphi} \right)\right] \rule{0mm}{4mm} \right\} \; ,
\end{align}
while the charge density and three-dimensional current density are:
\begin{align}
\rho_Q &= \frac{iq}{8\pi} \left( x_r \bar{\mathcal{E}}^r - \bar{x}_r \mathcal{E}^r \right) \; , \\
j^r_Q &= \frac{iq}{8\pi} \left( \varphi \bar{\mathcal{E}}^r - \bar{\varphi} \mathcal{E}^r \right) \; .
\end{align}

Similarly, the evolution equations for the EMP system now become:
\begin{align}
\partial_t \varphi &= \beta^r \partial_r\varphi - \frac{x_r}{\mathper{a}} \: \partial_r\alpha  - \frac{\alpha}{\mathper{a}} \left[\partial_rx_r +x_r \left(\frac{2}{r} - \frac{\partial_r \mathper{a}}{2\mathper{a}} + \frac{\partial_r \mathper{b}}{\mathper{b}} \right)\right] + \alpha \varphi K \notag \\
    & \hspace{4.3cm} + iq\alpha \left(-\phi\varphi + \frac{1}{\mathper{a}} a_rx_r -\frac{\mathper{a}}{m^2} E^r \mathcal{E}^r\right) \; ,
    \label{ev-varphi-3+1} \\
\partial_t \phi &=  \beta^r \partial_r\phi  - \frac{a_r}{\mathper{a}} \: \partial_r\alpha - \frac{\alpha}{\mathper{a}} \left[\partial_ra_r +a_r \left(\frac{2}{r} - \frac{\partial_r \mathper{a}}{2\mathper{a}} + \frac{\partial_r \mathper{b}}{\mathper{b}} \right)\right] + \alpha \phi K \; ,
\label{ev-phi-3+1} \\
\partial_tx_r &= \beta^r \partial_rx_r + x_r\partial_r\beta^r - \alpha (\mathper{a} \mathcal{E}^r + \partial_r\varphi) - \varphi\partial_r\alpha - iq\alpha \left( x_r \phi - a_r \varphi \right) \; ,
\label{ev-x-3+1} \\
\partial_ta_r &= \beta^r \partial_ra_r + a_r\partial_r\beta^r - \alpha (\mathper{a}E^r + \partial_r\phi) - \phi\partial_r\alpha \; ,
\label{ev-a-3+1} \\
\partial_t \mathcal{E}^r &= \beta^r\partial_r \mathcal{E}^r - \mathcal{E}^r\partial_r\beta^r + \alpha \left(K\mathcal{E}^r + \frac{m^2}{\mathper{a}} x_r\right) - iq\alpha\phi \mathcal{E}^r \; ,
\label{ev-Ep-3+1} \\
\partial_t E^r &= \beta^r\partial_r E^r - E^r\partial_r\beta^r + \alpha KE^r - 4\pi\alpha j_Q^r \; . \label{ev-Em-3+1}
\end{align}
Notice that we no longer have evolution equations for the magnetic fields since they vanish.
Similarly, the magnetic constraints trivialize too, and electric constraints become:
\begin{align}
\partial_r E^r + E^r \left(\frac{2}{r} + \frac{\partial_r \mathper{a}}{2\mathper{a}} + \frac{\partial_r \mathper{b}}{\mathper{b}} \right) &= 4\pi\rho_Q \; .
\label{cons-Em-3+1} \\
\partial_r \mathcal{E}^r + \mathcal{E}^r \left(\frac{2}{r} + \frac{\partial_r \mathper{a}}{2\mathper{a}} + \frac{\partial_r \mathper{b}}{\mathper{b}} \right) &= - m^2\varphi + iqa_r\mathcal{E}^r \; ,
\label{cons-Ep-3+1}
\end{align}


\section{Charged Proca stars}

We are looking for solutions for a static spacetime, which implies that the components of the metric do not depend on time and they are invariant with respect to time reflections. That means in particular that the shift vector must vanish, $\beta^r=0$.  Furthermore, we adopt the area radial coordinate by setting $\mathper{b}=1$, which ensures that the area of a sphere of constant radius is always $4\pi r^2$. These two conditions together reduce the metric to:
\begin{equation}
ds^2 = - \alpha(r)^2 dt^2 + \mathper{a}(r) dr^2 + r^2  d \Omega^2 \; ,
\end{equation}
Finally, since the spatial metric $\gamma_{ij}$ does not depend on time and we have $\beta^r=0$, this implies that the extrinsic curvature must also vanish, $K_{ij}=0$.

For the charged Proca star we propose the following ansatz for the Proca potentials:
\begin{equation} \label{ansatz-pot}
\varphi(t,r) = \varphi(r)e^{-i\omega t} \; , \qquad
x_r(t,r) = ix(r)e^{-i\omega t} \; ,
\end{equation}
with $\varphi(r)$ and $x(r)$ real, and where we have reused the symbols to keep the notation simple. When we formulate the potentials as in (\ref{ansatz-pot}), it is easy to show that we eliminate the time dependence of the quantities derived from the stress-energy tensor and the charge density, consistent with having a static spacetime. It is also easy to verify that $J^r=0$ and $j_Q^r=0$, and since $B^i=0$ we are faced with an electrostatics problem. The Maxwell potentials then take the simple form:
\begin{equation}
\phi(t,r) = \phi(r) \; , \qquad
a_r(t,r) = 0 \; .
\end{equation}

Let us now consider the EMP evolution equations (\ref{ev-varphi-3+1}-\ref{cons-Em-3+1}). First we consider the evolution equations for the electric fields. From the Maxwell equation (\ref{ev-Em-3+1}) we deduce that $E^r = E(r)$, while the Proca equation (\ref{ev-Ep-3+1}) results in $\mathcal{E}^r = \mathcal{E}(r)e^{-i\omega t}$ with: 
\begin{equation} \label{eq-Ep-analitic}
\mathcal{E} = - \frac{\alpha m^2 x}{\mathper{a}(\omega - q\alpha\phi)} \; ,
\end{equation}
Next we consider the evolution equations for the potentials, we obtain:
\begin{align}
\varphi' &= - \mathper{a} \mathcal{E} - \frac{(\omega - q\alpha\phi)x} {\alpha} - \frac{\varphi \alpha'} {\alpha} \; ,
\label{eq-rad-varphi-0} \\
x' &= \frac{(\omega - q\alpha\phi)\varphi \mathper{a}}{\alpha} - x \left(\frac{2}{r} - \frac{\mathper{a}'}{2\mathper{a}} + \frac{\alpha'}{\alpha}\right) - q\frac{\mathper{a}^2E \mathcal{E}}{m^2} \; ,
\label{eq-rad-x-0} \\
\phi' &= -\mathper{a} E - \frac{\phi\alpha'}{\alpha} \; ,
\label{eq-rad-phi-0}
\end{align}
where the prime now denotes derivatives with respect to $r$. These equations are found as follows: (\ref{eq-rad-varphi-0}) from (\ref{ev-x-3+1}), (\ref{eq-rad-x-0}) from (\ref{ev-varphi-3+1}), and (\ref{eq-rad-phi-0}) from (\ref{ev-a-3+1}), while (\ref{ev-phi-3+1}) is now satisfied trivially. Finally, by considering the electric constraints (\ref{cons-Em-3+1}) and (\ref{cons-Ep-3+1}), we arrive respectively at:
\begin{align}
E' = - qx\mathcal{E} - E \left(\frac{2}{r} + \frac{\mathper{a}'}{2\mathper{a}}\right) \; . \label{eq-rad-Em-0} \\
\mathcal{E}' = -m^2\varphi - \mathcal{E} \left(\frac{2}{r} + \frac{\mathper{a}'}{2\mathper{a}}\right) \; ,
\label{eq-rad-Ep-0}
\end{align}
Note that in (\ref{eq-Ep-analitic}) we have already found an analytical relationship between $\mathcal{E}$ and some other variables, which is completely consistent with (\ref{eq-rad-Ep-0}), therefore this equation is unnecessary in our case (but notice that~(\ref{eq-Ep-analitic}) will no longer be valid for a dynamical time evolution).

To complete the system we still need the equations for the radial metric and the lapse function. These are, respectively:
\begin{align}
\mathper{a}' &= \mathper{a} \left[\frac{1}{r}(1-\mathper{a}) + 8\pi r \mathper{a} \rho\right] \; , \label{eq-rad-a} \\
\alpha' &= \alpha \left[\frac{1}{2r}(\mathper{a}-1) + 4\pi r \mathper{a} S^r_{\phantom{r}r}\right] \: . \label{eq-rad-alpha}
\end{align}
The equation for the radial metric $\mathper{a}(r)$ can be derived directly from the Hamiltonian constraint, which in spherical symmetry and with the choices we have made reduces to~(\ref{eq-rad-a}). For the lapse function $\alpha(r)$ we use the so-called polar slicing condition that comes from having chosen the areal radius, the ADM equations for the extrinsic curvature then imply~(\ref{eq-rad-alpha}). Next, we use the metric and lapse equations to eliminate the radial derivatives of $\mathper{a}$ and $\alpha$ from equations (\ref{eq-rad-varphi-0}-\ref{eq-rad-Em-0}). We find:
\begin{align}
\psi' &= - \mathper{a} \mathcal{E} - \frac{(\omega - q\vartheta)x} {\alpha} \; ,
\label{eq-rad-psi} \\
x' &= \frac{(\omega - q\vartheta) \psi \mathper{a}}{\alpha^2} - x \left[ \frac{1}{r} \left( \mathper{a} + 1 \right) + 4 \pi r\mathper{a} \left( S^r_{\phantom{r}r} - \rho \right) \right] - q \: \frac{\mathper{a}^2 E \mathcal{E}}{m^2} \; ,
\label{eq-rad-x} \\
\vartheta' &= -\alpha \mathper{a}E, \label{eq-rad-vartheta} \\
E' &= - q x \mathcal{E} - E \left[ \frac{1}{2r}(5-\mathper{a}) + 4 \pi r \mathper{a} \rho \right] \; ,
\label{eq-rad-E}
\end{align}
where we have also introduced the variable changes $\psi := \alpha \varphi$ and $\vartheta := \alpha \phi$ in order to simplify the expressions.

Finally, for the matter terms we find:
\begin{align}
\rho = + \frac{1}{8\pi} \left[ \mathper{a} \mathcal{E}^2 + m^2 \left(\varphi^2 + \frac{x^2}{\mathper{a}}\right) + \mathper{a} E^2\right] \; ,
\label{rho-epc} \\
S^r_{\phantom{r}r} = -\frac{1}{8\pi} \left[ \mathper{a} \mathcal{E}^2 - m^2 \left(\varphi^2 + \frac{x^2}{\mathper{a}}\right) + \mathper{a} E^2 \right] \; .
\label{s-epc}
\end{align}

Equations (\ref{eq-rad-a}-\ref{eq-rad-E}), together with expressions (\ref{eq-Ep-analitic}), (\ref{rho-epc}) and (\ref{s-epc}), form the complete system of radial equations for the charged Proca star (CPS). This is a system of six coupled first order non-linear ordinary differential equations in $r$ for the six dependent variables $\{ \mathper{a}, \alpha, \psi, x,\vartheta, E\}$. Notice also that when we take $q=0$ the system reduces to the well-known system of four radial equations of an uncharged Proca star.


\section{Boundary conditions, rescaling and gauge transformation}

To solve the system of radial equations described above it is necessary to set boundary conditions. We start with the conditions at infinity, where we must recover Minkowski spacetime, and set up the potentials and fields associated with matter to be zero, since we want to model a compact object. We then ask for:
\begin{equation} \label{cond-inf}
\alpha, \mathper{a} \rightarrow 1 \; , \qquad
\varphi, x, \phi, E \rightarrow 0 \; .
\end{equation}
We can be somewhat more precise.  If we assume that far away we have Minkowski spacetime so that $\alpha=\mathper{a}=1$, and vanishing Maxwell fields $\phi=E=0$, equations (\ref{eq-rad-psi}) and (\ref{eq-rad-x}) then simplify to:
\begin{equation}
\varphi' = \omega x \left( \frac{m^2}{\omega^2} - 1\right) \; , \qquad
x' = \omega \varphi \; ,
\end{equation}
which is a simple system and can be solved analytically to find:
\begin{equation} \label{dec-exp}
\varphi,x \rightarrow \exp \left(\pm \sqrt{m^2 - \omega^2} \right) \; .
\end{equation}
For this expression to be real, it is necessary that the condition $\omega \leq m$ is fulfilled. Moreover, since the boundary condition imposes that $\varphi$ must decay to zero at infinity, this is only possible if the coefficient of the increasing exponential vanishes. This condition is only satisfied for certain values of $\omega$, which leads to the conclusion that the values of $\omega$ must have a discrete spectrum.  In other words, we have to solve an eigenvalue problem.

We now set the conditions at the origin for the charged Proca star. These must be:
\begin{equation} \label{cond-orig}
\mathper{a}(0)=1, \quad \alpha(0)=\alpha_0, \quad \varphi(0)=\varphi_0, \quad x(0)=0, \quad \phi(0)=\phi_0, \quad E(0)=0,
\end{equation}
with $\alpha_0$, $\phi_0$ and $\varphi_0$ positive real constants. Asking for $\mathper{a}(0)=1$ is required in order to guarantee that the spacetime is locally flat at the origin, while we must have $x(0)=E(0)=0$ because they correspond to the radial components of vectors in a spherically symmetrical spacetime. Since we do not have prior information determining the values of the $\alpha$ and $\phi$ at the origin, for the numerical integration we provisionally take $\alpha_0=1$ and $\phi_0=0$ (but see below). Finally, $\varphi(0)=\varphi_0$ is our free parameter.

Replacing these conditions in the r adial equations of the scalar potentials $\varphi$ and $\phi$, we deduce their derivatives at the origin, while by regularity the derivatives of the geometric variables $\mathper{a}$ and $\alpha$ must be zero. Summarizing we find:
\begin{equation}
\mathper{a}'(0)=0 \; , \quad
\alpha'(0)=0 \; , \quad
\varphi'(0)=0 \; , \quad
\phi'(0)=0 \; .
\end{equation}

For the radial derivatives of $x$ and $E$ at the origin we must be more careful. The radial equation for $x$, equation~\eqref{eq-rad-x} has a term that varies as $x/r$, suggesting that close to the origin $x$ should be proportional to $r$. That is, we can express $x = k_x r$, with $k_x$ some constant. Substituting this into the radial equation and using the conditions above we immediately obtain:
\begin{equation}
x'(0) = k_x = \frac{\omega \varphi_0}{3} \; ,
\end{equation}
where we have already taken $\alpha_0=1$ and $\phi_0=0$.

Let us now consider the radial equation for $E$, which near the origin simply reduces to $E' = -2E/r$. If we now take $E = k_E r$ we immediately find $k_E=0$.  In fact, from the structure of the equation it is not difficult to show that close to the origin we must in fact have $E \sim r^3$, so that:
\begin{equation}
E'(0) = 0 \; ,
\end{equation}

We can also give some more detail. The radial equation for $\mathper{a}(r)$ contains a term that goes as $(1-\mathper{a})/r$. This suggests that, near the origin, the radial metric must take the form $\mathper{a} = 1 + k_a r^2$. Plugging this into the radial equation we find $k_{\mathper{a}} = 8 \pi \rho_0 / 3$,
with $\rho_0 \equiv \rho(0) = m^2 \varphi_0^2 / 8\pi$, so that finally:
\begin{equation}
k_{\mathper{a}} = \frac{1}{3} \: m^2 \varphi_0^2 \; .
\end{equation}

There is a couple of final details we must address. First, we must note that taking $\alpha_0=1$ is not compatible with asking for $\alpha=1$ at infinity. Fortunately, this problem is easily overcome since the lapse equation is linear, which allows us to rescale $\alpha$ at the end of the process to correct its asymptotic value. However, when we do this we must also rescale the frequency and the scalar potentials $\phi$ and $\varphi$ since the whole system of equations is invariant under the transformation:
\begin{equation} \label{reescal}
\alpha \rightarrow \alpha/C \; , \qquad
\omega \rightarrow \omega/C \; , \qquad
\psi \rightarrow \psi/C \; , \qquad
\vartheta \rightarrow \vartheta/C \; ,
\end{equation}
with $C$ an arbitrary constant constant. In practice we take $C=\alpha_\infty$, where $\alpha_\infty$ is the asymptotic value of the lapse resulting from solving the system. The value of $\alpha_0$ is obtained by extrapolation assuming that for large $r$ the
lapse behaves as $\alpha = \alpha_\infty + k/r$, with $k$ some constant. From this we find that $\alpha_\infty = \alpha + r \alpha'$, where $\alpha'(r \rightarrow \infty)$ can be determined by numerical differentiation at large radius.

Similarly, taking $\phi_0=0$ is not compatible with asking for $\phi$ to go to zero at infinity. To fix this, once we have done the numerical integration of the system we must do a gauge transformation of the form:
\begin{equation}
\vartheta \rightarrow \vartheta - \partial_t \theta \: , \qquad
\psi \rightarrow \psi e^{-iq\theta} \; , \qquad
x \rightarrow x e^{-iq\theta} \; , \qquad
\omega \rightarrow \omega + q \partial_t \theta \; ,
\end{equation}
with $\theta=\vartheta_\infty t$, and where now $\vartheta_\infty$ is the asymptotic value of $\vartheta$ at infinity obtained after the rescaling given in (\ref{reescal}). The gauge transformation is evaluated at $t=0$, so that in practice we only have:
\begin{equation}
\label{gauge}
\vartheta \rightarrow \vartheta - \vartheta_0 \: , \qquad
\omega \rightarrow \omega + q \vartheta_0 \; .
\end{equation}

Notice that both the rescaling and the gauge transformation involve changes in the frequency $\omega$. Thus, for a given value of $q$ and $\varphi_0$ we will have three distinct values for the frequency: An initial value $\omega_1$ that comes from solving the eigenvalue problem under the provisional boundary conditions at the origin, a second value $\omega_2$ obtained after rescaling the lapse, and a final value $\omega$ obtained after gauge transformation. Rescaling and gauge transformation, applied in that order, eventually lead us to the desired solution for the families of charged Proca stars. In the numerical results of the following section we will only report the final transformed values.


\section{Total mass, charge, and binding energy}

It is well known that when we work with the areal radius in spherical symmetry, the total mass $M$ is simply given by the integral of the energy density over a flat volume element (this result can be derived directly from the Hamiltonian constraint for a static spherically symmetric spacetime):
\begin{equation}
M = 4 \pi \int_0^\infty \rho r^2 dr \; .
\end{equation}
which for charged configurations converges as $1/r$ since $\rho$, as given in equation~(\ref{rho-epc}), decays as $1/r^2$ due to the presence of the electric field $E$. On the other hand, the total electric charge $Q$ is given by:
\begin{equation}
Q := 4\pi \int_0^\infty \rho_Q \mathper{a}^{1/2} r^2dr \; ,
\end{equation}
which converges exponentially, since the charge density is given by:
\begin{equation}
\rho_Q = -\frac{q}{4\pi} \: x \mathcal{E} \; ,
\end{equation}
and both $\mathcal{E}$ and $x$ decay exponentially. Do notice that, in constrast with the mass integral above, the charge integral is in fact done over the curved (physical) volume element.
We can also define a total number of particles (bosons) $N$ which is related to the total charge $Q$ simply by $N=Q/q$.  Notice that this $N$ can be defined even for the case with $q=0$ since the value of $q$ cancels out.

Since the total mass integral converges quite slowly with distance, we use instead a different approach to calculate the total mass of a given configuration. The mass can also be found by assuming that at large distances the spacetime reduces to the Reissner–Nordström (RN) solution. This has to be so since, as mentioned above, the Proca field decays exponentially and the RN solution it is the only static solution with spherical symmetry for an electro-vacuum spacetime. This implies that in then electrovac region the metric function $\mathper{a}(r)$ must converge to the characteristic form of the radial component of the Reissner-Nordström metric:
\begin{equation}
a(r) \rightarrow \left( 1 - \dfrac{2M}{r} + \dfrac{Q^2}{r^2} \right)^{-1} \; .
\end{equation}
Solving for M and taking the limit as $r$ goes to infinity, we find:
\begin{equation} \label{masa-rn}
M = \lim_{r \rightarrow \infty} \left[\frac{r}{2} \left(1+ \frac{Q^2}{r^2} - \frac{1}{\mathper{a}} \right) \right] \; .
\end{equation}
Notice that, since $Q$ converges exponentially, this expression should also converge to the total mass very fast as we increase the value of $r$.

\vspace{5mm}

We can use the total mass $M$ and the total number of particles $N$ to define a binding energy as:
\begin{equation} \label{eq-Eb}
E_B:= M - mN = M - \left( m/q \right) Q \; .
\end{equation}
This binding energy corresponds to the difference between the total mass-energy of the star $M$, and the total rest mass the bosonic particles $mN$, so it effectively measures the difference between the kinetic and potential energy contributions.  Stars with $E_B<0$ should therefore be gravitationally bound, while those with $E_B>0$ should be unbound.


\section{Numerical Results} 

In this section, we present the results obtained by performing the numerical integration of the system of radial equations derived above. For a given value of the Proca scalar potential $\varphi_0$ at the origin, we solve the eigenvalue problem by using a shooting algorithm to find the correct frequency $\omega$ that allows us to satisfy the boundary conditions at large radius, combined with a fourth-order Runge-Kutta method for the integration of the radial differential equations. In all reported cases, the radial grid spacing was set to $\Delta r = 0.01$, with a minimum of $N_r=4000$ grid points along the radial direction (setting our boundaries at r=40.0).  We have also checked that increasing resolution and/or moving the boundaries further away does not affect our results in a significant way.

For simplicity, we choose the mass parameter $m=1$. However, this value can be later rescaled to any arbitrary value since the system of equations is also invariant under the following transformation:
\begin{equation}
m \rightarrow \lambda m \; , \qquad 
q \rightarrow \lambda q \; , \qquad
\omega \rightarrow \lambda \omega \; , \qquad
r \rightarrow r/\lambda \; , \qquad 
E \rightarrow \lambda E \; , \qquad
\mathcal{E} \rightarrow \lambda \mathcal{E} \; ,
\end{equation}
with $\{\varphi, x, \phi, \mathper{a}, \alpha\}$ unchanged.

It is important to mention at this point that there is a critical value of the charge $q_c$ above which one would not expect to find solutions to our system of equations, which in our units corresponds to $q_c=m$ (or $q_c=1$ for $m=1$).  This critical value comes from the fact that for charged configurations there is a repulsive Coulomb force that acts against gravity.  For $q>q_c$ this repulsive force is stronger than gravity, at least in the newtonian case, so one would not expect stationary solutions to exist. In studying charged boson stars such a critical value of the charge has been shown to play a very important role.  However, it has been reported in~\cite{Pugliese_2013,Lopez:2023phk} that stationary solutions for charged boson stars can in fact be found for values of $q$ slightly above the critical value. We will show below that this is also the case for charged Proca stars, and supercritical solutions also exist for $q>q_c$, but only for a very narrow range of values of both $q$ and $\varphi_0$.


\subsection{Families of solutions for different values of the charge}

For an electric charge $q$ (first degree of freedom) we have a specific CPS family, where each star is characterized by a value of the central Proca scalar potential $\varphi_0$ (second degree of freedom) and has a set of well-defined properties. In this section, we show our main results for the different CPS families. In previous work \cite{Brito:2015yfh,SalazarLandea:2016bys}, instead of using the potential $\varphi_0$ as a parameter as we do here, it has been preferred to use the component $X_0$ of the Proca potential at the origin. We recall that these two quantities are related by $X_0 = -\alpha\varphi$. From this we define:
\begin{equation}
\psi_0 := -X_0(r=0) = \alpha_0 \varphi_0 \; , 
\end{equation}
where $\alpha_0$ is the value of the lapse at the origin. In practice, the above relation is a simple rescaling of the potential. In order to compare our results with previous ones, in this section we will use $\psi_0$ instead of $\varphi_0$ as a free parameter and call it the ``central potential".

\subsubsection{Frequency}

\begin{figure}[h]
\centering
\includegraphics[width=0.8\textwidth]{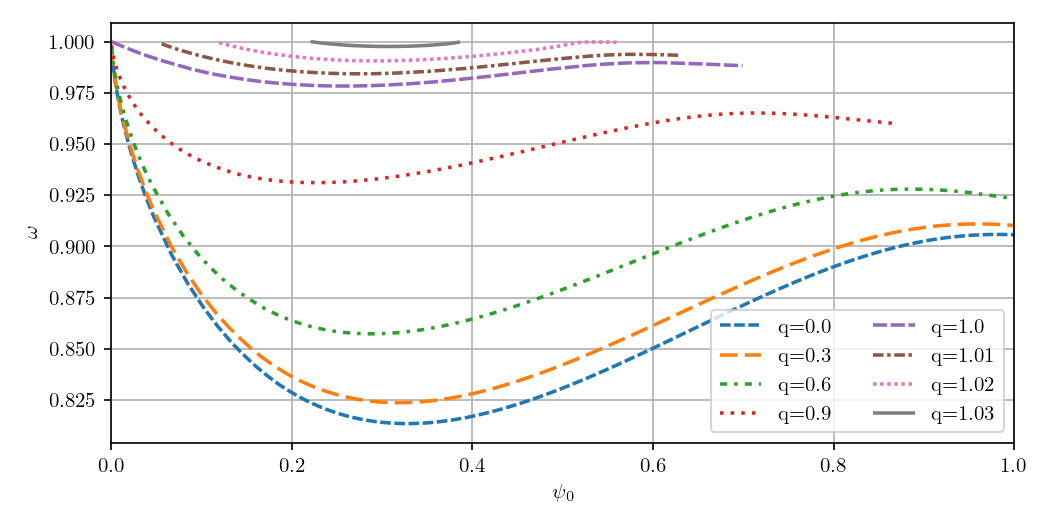}
\caption{\textit{Frequency $\omega$ as a function of the central value of scalar potential $\psi_0$ for families of charged Proca stars with different values of the charge parameter $q$.}}
\label{omega-vs-phi0}
\end{figure}

Figure~\ref{omega-vs-phi0} shows how the frequency $\omega$ varies as a function of the central potential $\psi_0$, for different values of the charge parameter $q$. This graph summarizes the entire set of allowed CPS families. For stars with $q\leq1$, we can find solutions for any positive value of $\psi_0$. However, because the numerical integration of the equations was performed only up to certain values of $\psi_0$, the curves in the plots for $q=0.9$ and $q=1.0$ appear truncated. In contrast, for supercritical charges such that $q>1$ the value of the central potential $\psi_0$ is no longer arbitrary, and CPS solutions are only found for certain specific intervals. Remember that, as discussed above, the boundary conditions imply that our solutions must be such that $\omega \leq m = 1$, so no solutions can exist for $\omega>1$. In particular, for the supercritical solutions there is both a lower and an upper bound for $\psi_0$, and solutions can not be found outside of this interval.  We have set $q=1.04$ as an upper bound because we have found no solutions above that value.

It is well known that for uncharged Proca stars the value of $\omega$ oscillates slowly as $\psi_0$ increases, and converges to $\omega \sim 0.891$~\cite{Brito:2015pxa}. This is most clearly seen in the $M$ vs $\omega$ curve in Figure~\ref{masa-vs-frecuencia}, where the solution tends to an attractor with fixed mass and frequency values. Our results show that CPS's with $q\leq1$ seem to have a similar behavior but with larger frequencies, with each curve oscillating and apparently converging to a value that depends on the charge. On the other hand, for supercritical charges, these oscillations are truncated at a certain value of $\psi_0$. In all cases, there is a minimum value for the frequency that depends on the charge. For subcritical values of the charge, the position of this minimum moves to the left as the charge increases.  However, this behavior is reversed when supercritical charged values are considered, with the position of the minimum moving to the right.

\subsubsection{Effective radius}

\begin{figure}[h]
\centering
\includegraphics[width=0.8\textwidth]{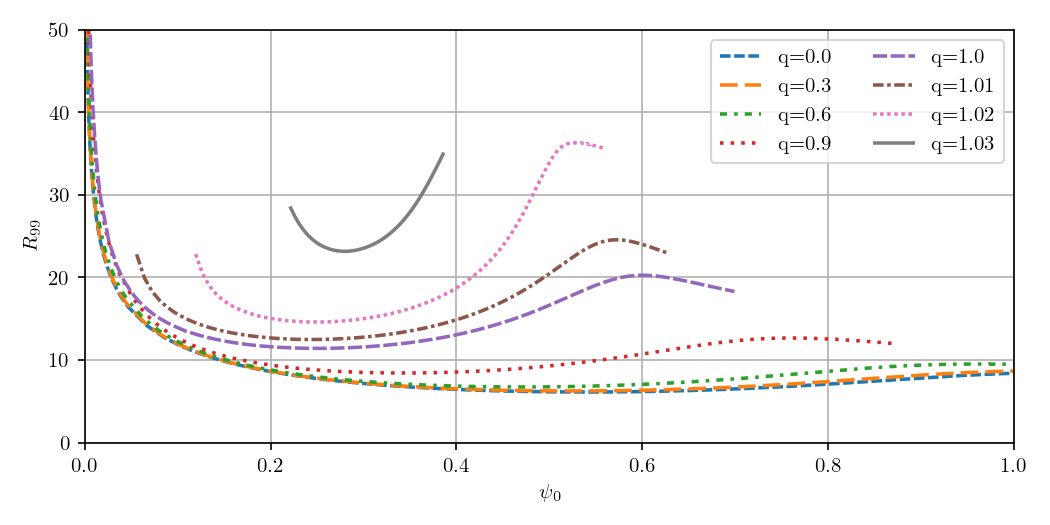}
\caption{\textit{Effective radius $R_{99}$ as a function of the central potential $\psi_0$ for the different CPS families.}}
\label{radio-vs-phi0}
\end{figure}

CPS do not have a precise external surface since the matter fields decay exponentially. We therefore introduce the concept of an``effective radius" $R_{99}$, which is operationally defined as the distance within which $99\%$ of the total number of particles $N$ of the star is contained.  Notice that this effective radius is often defined as that which contains $99\%$ of the mass, but in the case of charged configurations this definition is not ideal since the electric field contributes to the mass and decays very slowly, so here we prefer to define it as the radius containing $99\%$ of the particles. Figure~\ref{radio-vs-phi0} shows how $R_{99}$ varies as a function of $\psi_0$ for the different values of the electric charge $q$.

For $q \leq 1$ the effective radius of the star can be very large for small values of $\psi_0$, indicating a highly dispersed structure. However, as we consider stars with a larger central potential, $R_{99}$ decreases dramatically until it reaches a minimum. Beyond this, the effective radius shows attenuated oscillations.

For supercritical charges $q>1$, the behavior of the curves reveals distinctive features. On the one hand, since there is a lower bound for $\psi_0$, solutions with arbitrarily large radii do not exist. Furthermore, the effective radius for supercritical charges also exhibits a minimum as a function of $\psi_0$, which shifts more and more to the right as the charge increases.

\subsubsection{Total mass and charge}

\begin{figure}[h]
\centering
\includegraphics[width=0.9\textwidth]{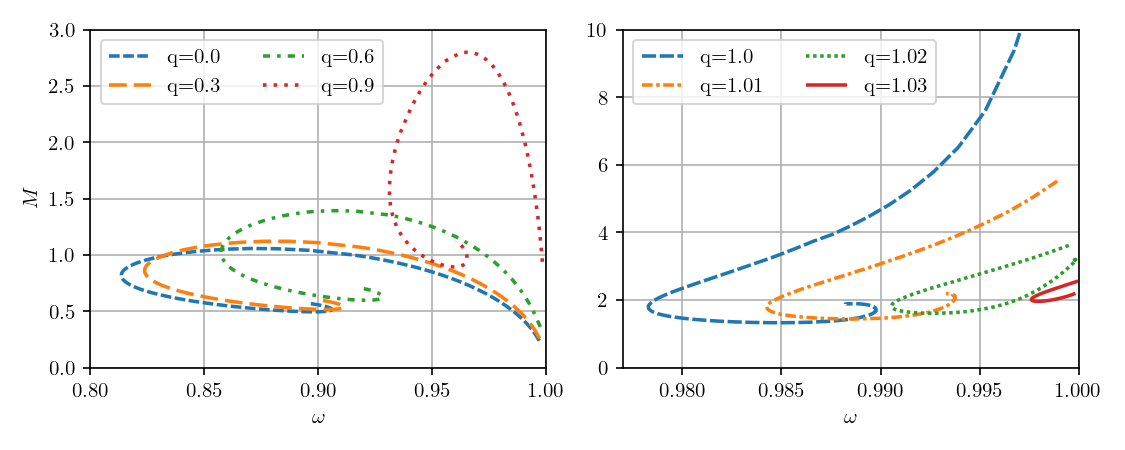}
\caption{\textit{Total mass $M$ as a function of frequency $\omega$ for different CPS families. The left panel corresponds to families with subcritical charges and the right panel to families with supercritical charges (including the critical charge $q=1$).}}
\label{masa-vs-frecuencia}
\end{figure}

Figure~\ref{masa-vs-frecuencia} shows a plot of the total mass $M$ in terms of the frequency $\omega$ for the different families of solutions. For clarity, we show subcritical solutions on the left panel and supercritical solutions on the right.  When the charge is subcritical, the curves take on a spiral shape that grows counterclockwise as $\psi_0$ increases, and then converges towards a specific attraction point. This suggests that for very large values of $\psi_0$ the CPS reaches a state in which its properties tend toward asymptotic stability, where mass and frequency stabilize at a predictable end point. Interestingly, the coordinates of this attractor point are higher in both frequency and mass for families of stars with a higher charge. This means that, as the charge increases, both the frequency and the mass tend to converge towards progressively higher values, as previously observed in Figure~\ref{omega-vs-phi0}.

When the charge is supercritical, the curves do not show the typical spiral behavior observed for the subcritical cases, although they are somewhat similar. Specifically, the curve corresponding to the critical value of charge $q=1$ is distinguished by not being bounded from above. At one extreme, the mass tends to infinity as the frequency approaches $\omega=1$, while at the other the curve presents an attractor point, as occurs in families of stars with subcritical charge. In strictly supercritical charge cases, the attractor is not formed, since the central potential $\psi_0$ cannot be arbitrarily large.

\begin{figure}[h]
\centering
\includegraphics[width=0.9\textwidth]{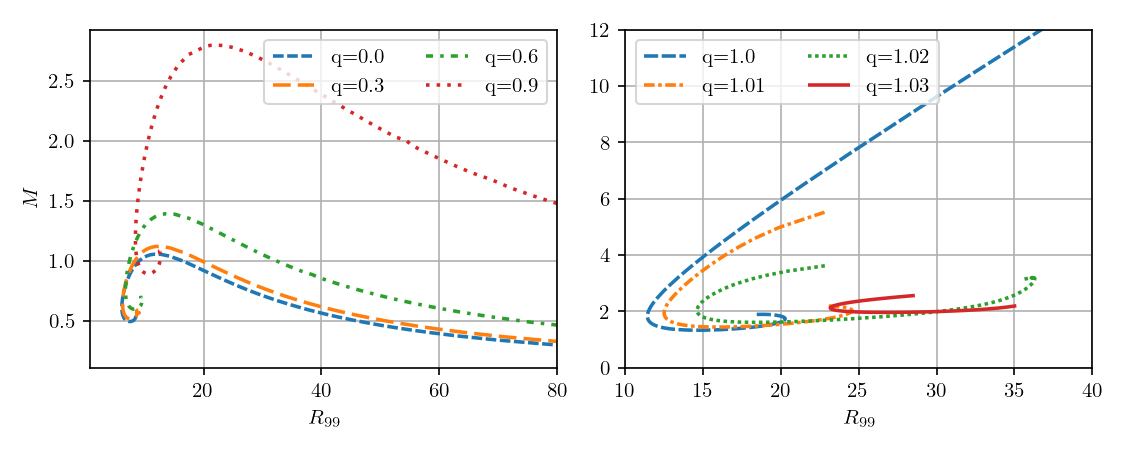}
\caption{\textit{Total mass $M$ as a function of the effective radius $R_{99}$ for different CPS families. The left panel corresponds to families with subcritical charges and the right panel to families with supercritical charges (including the critical charge $q=1$).}}
\label{masa-vs-radio}
\end{figure}

Figure~\ref{masa-vs-radio} shows the relationship between mass and effective radius for different families of CPS. When the charge is subcritical (left panel), the curves present a slight concavity; until at a certain point they begin to adopt a spiral profile that converges towards an attractor. On the other hand, when the charge is supercritical (right panel), the curves have a very different profile. For the particular case of $q=1$, for small values of $\psi_0$ the relationship between mass and effective radius is practically linear, and this pattern persists even for effective radii very close to the minimum radius of the star. From a certain point, the curve begins to exhibit a spiral behavior, heading towards an attractor similar to that observed in families with subcritical charges.

\begin{figure}[h]
\centering
\includegraphics[width=0.9\textwidth]{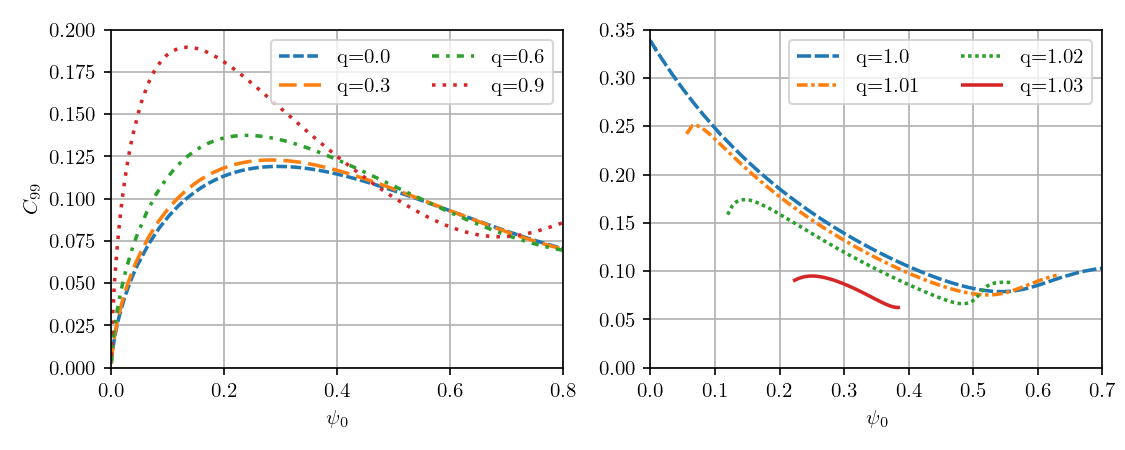}
\caption{\textit{Compactness $C_{99}$ as a function of the central potential $\psi_0$ for different CPS families. The left panel corresponds to the subcritical charge regime. The right panel to the supercritical charge regim (including the critical charge $q=1)$.}}
\label{compacticidad-vs-phi0}
\end{figure}

Figure~\ref{compacticidad-vs-phi0} shows the effective compactness $C_{99} := M/R_{99}$ of the different families of solutions as a function of the central potential $\psi_0$. The profiles of the curves of families with subcritical (left) and supercritical (right) charges are well differentiated. The former exhibit a maximum compactness for some value of $\psi_0$ that becomes larger and moves to the left as $q$ increases. The latter are decreasing in most of the allowed interval, except at the extremes. When the charge is critical, the compactness reaches the maximum value $C_{99} \simeq 0.34$ as $\psi_0 \rightarrow 0$ (compare this with the compactness of a Schwarzschild black hole $C=0.5$, and that of a maximally charged Reissner-Nordstrom black hole $C=1$).

\subsubsection{Binding energy}

\begin{figure}[h]
\centering
\includegraphics[width=0.8\textwidth]{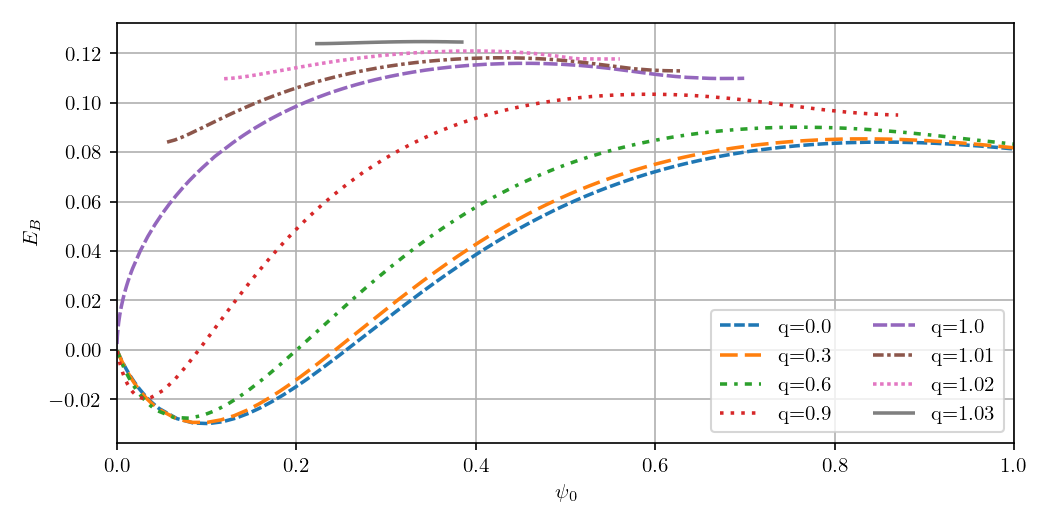}
\caption{\textit{Binding energy $E_B$ as a function of the central potential $\psi_0$ for different CPS families.}}
\label{energ-amarre-vs-phi0}
\end{figure}

Figure~\ref{energ-amarre-vs-phi0} shows the binding energy $E_B$ as a function of the central potential for different CPS families. For families with subcritical charges, solutions can be either gravitationally bound ($E_B<0$) or unbound ($E_B>0$).  The higher the value of the charge $q$, the smaller the interval of bound solutions becomes. Notice also that for all families the minimum value of the binding energy $E_B$ always corresponds to the maximum value of the total mass $M$. On the other hand, for critical and supercritical charges there exist only unbound solutions. 

The study of the binding energy of stars is essential to understand their possible dynamics. Dynamical stability analysis carried out in~\cite{Alcubierre:2019qnh,Brito:2015pxa,SanchisGual:2017,Brito:2024,Pugliese_2013,Kain_2021,SalazarLandea:2016bys}, both for boson stars and for Proca stars, show curves with profiles similar to those we have obtained here. They have found that gravitationally unbound stars are always unstable to perturbations, and when they are perturbed they either collapse to a black hole or disperse to infinity. Furthermore, in the branch with $E_B<0$ two regions can be distinguished, one where the solutions are stable (to the left of the minimum) and the other where they are unstable (to the right of the minimum). The latter can migrate towards a star in the stable branch or collapse to a black hole when perturbed. We speculate that our solutions will have a similar behavior, and we are currently performing numerical dynamical simulations to study this that will be reported elsewhere.

The minimum values of $E_B$ for each curve in Figure~\ref{energ-amarre-vs-phi0}, which correspond to the maximum mass $M$, as well as the points for which the binding energy is zero, are summarized in tables~\ref{tab:datos1} and \ref{tab:datos2}.

\begin{table}[h]
    \centering
    \begin{minipage}{0.52\textwidth}
        \centering
        \begin{tabular}{ccccccccc}
        \hline
            $q$ & $\psi_0$ & $\omega$ & $Q$ & $N$ & $M$ & $R_{99}$ & $C_{99}$ & $E_B$ \\
            \hline
            0.0 & 0.098 & 0.874 & 0.000 & 1.088 & 1.058 & 12.097 & 0.087 & -0.030 \\
            0.1 & 0.097 & 0.874 & 0.109 & 1.095 & 1.065 & 12.105 & 0.088 & -0.030 \\
            0.2 & 0.094 & 0.878 & 0.223 & 1.116 & 1.086 & 12.368 & 0.088 & -0.030 \\
            0.3 & 0.092 & 0.882 & 0.346 & 1.153 & 1.123 & 12.523 & 0.090 & -0.030 \\
            0.4 & 0.089 & 0.888 & 0.484 & 1.210 & 1.181 & 12.846 & 0.092 & -0.029 \\
            0.5 & 0.083 & 0.897 & 0.647 & 1.295 & 1.266 & 13.331 & 0.095 & -0.029 \\
            0.6 & 0.076 & 0.908 & 0.853 & 1.422 & 1.395 & 14.082 & 0.099 & -0.028 \\
            0.7 & 0.066 & 0.923 & 1.138 & 1.625 & 1.599 & 15.329 & 0.104 & -0.026 \\
            0.8 & 0.052 & 0.941 & 1.584 & 1.981 & 1.957 & 17.427 & 0.112 & -0.024 \\
            0.9 & 0.032 & 0.965 & 2.538 & 2.820 & 2.800 & 22.119 & 0.127 & -0.020 \\
        \hline
        \end{tabular}
        \caption{Limit values for a stable CPS: When the binding energy is minimum, or the total mass is maximum.}
        \label{tab:datos1}
    \end{minipage}
    \hfill
    \begin{minipage}{0.47\textwidth}
        \centering
        \begin{tabular}{cccccccc}
            \hline
            $q$ & $\psi_0$ & $\omega$ & $Q$ & $N$ & $M$ & $R_{99}$ & $C_{99}$ \\
            \hline
            0.0 & 0.256 & 0.818 & 0.000 & 0.909 & 0.909 & 7.687 & 0.118 \\
            0.1 & 0.254 & 0.819 & 0.091 & 0.914 & 0.914 & 7.711 & 0.119 \\
            0.2 & 0.250 & 0.823 & 0.186 & 0.932 & 0.932 & 7.782 & 0.120 \\
            0.3 & 0.243 & 0.829 & 0.289 & 0.964 & 0.964 & 7.907 & 0.122 \\
            0.4 & 0.233 & 0.837 & 0.405 & 1.013 & 1.013 & 8.100 & 0.125 \\
            0.5 & 0.219 & 0.849 & 0.543 & 1.087 & 1.087 & 8.384 & 0.130 \\
            0.6 & 0.200 & 0.864 & 0.718 & 1.197 & 1.197 & 8.803 & 0.136 \\
            0.7 & 0.175 & 0.883 & 0.961 & 1.372 & 1.372 & 9.456 & 0.145 \\
            0.8 & 0.141 & 0.909 & 1.347 & 1.684 & 1.684 & 10.597 & 0.159 \\
            0.9 & 0.091 & 0.944 & 2.177 & 2.419 & 2.419 & 13.265 & 0.182 \\
            \hline
        \end{tabular}
        \caption{Limit values for a gravitationally bound CPS: When the binding energy is zero.}
        \label{tab:datos2}
    \end{minipage}
\end{table}

\begin{figure}[h]
    \centering
    \includegraphics[width=0.9\textwidth]{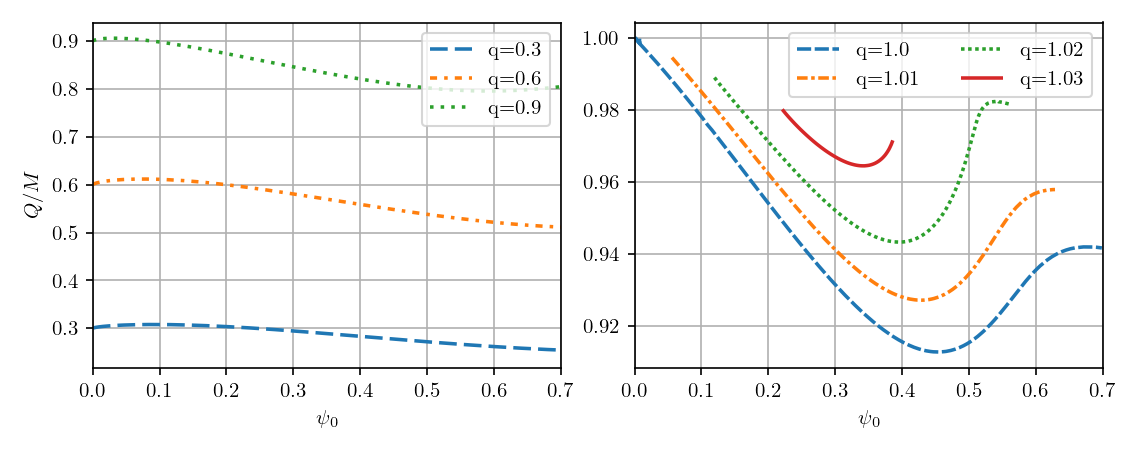}
    \caption{\textit{Charge-to-mass ratio $Q/M$ as a function of the central potential $\psi_0$ for different CPS families. The left panel corresponds to the subcritical charge regime. The right panel to the supercritical charge regime, including the critical charge.}}
    \label{razon-q/m-vs-phi0}
\end{figure}

If we factor out $M$ in equation~(\ref{eq-Eb}) we find another interesting way of writing the binding energy:
\begin{equation}
E_B = M \left[1- \left(\dfrac{m}{q}\right) \left(\dfrac{Q}{M}\right)\right] \; .
\end{equation}
This shows that for gravitationally bound solutions we must have $Q/M>q/m$, while for unbound solutions we must have $Q/M<q/m$. Figure~\ref{razon-q/m-vs-phi0} reveals a curious fact about the ratio $Q/M$. In the left panel we show the case of subcritical charges, where $Q<M$ always holds.  Furthermore, for small values of the central potential $\psi_0$ we also have $Q/M>q/m$, corresponding to gravitationally bound solutions. The right panel shows the case of supercritical charges.  We can also notice that $Q/M<1$ always holds, which is interesting since it means that even when $q/m>1$, the total charge $Q$ of the star will never be greater than its total mass $M$. Notice also that $Q=M$ is only reached for the critical case $q=1$ (remember that we are taking $m=1$) in the limit when $\psi_0$ goes to cero.

At this point, we must mention the fact that we were able to reproduce the results of Landea and García about CPS's reported in~\cite{SalazarLandea:2016bys} only partially. In their work they considered only the family corresponding to a charge $\tilde{q} = 0.5$ ($q=\sqrt{2}/2 \approx 0.7071$ in our convention).  In our work, we find for this family a maximum mass $M_{max} \simeq 1.62$, and a maximum number of particles $N_{max} \simeq 1.643$, which are slightly higher than those reported in~\cite{SalazarLandea:2016bys}: $M_{max} \simeq 1.56$ and $N_{max} \simeq 1.64$. We believe that this difference probably comes from the fact that they took the outer boundary of their integration domain too close, and also that their resolution was lower than ours. Interestingly, their value for the maximum total number of particles is very similar to our own, whereas their value for the maximum total mass is lower, reinforcing our supposition that they probably did not integrate the mass contribution far enough. Also, the curve they show for this family of solutions is not completely smooth and has a clear kink, again probably because of lack of resolution, whereas our results do not present this feature.

\begin{figure}[h]
\centering
\includegraphics[width=0.7\textwidth]{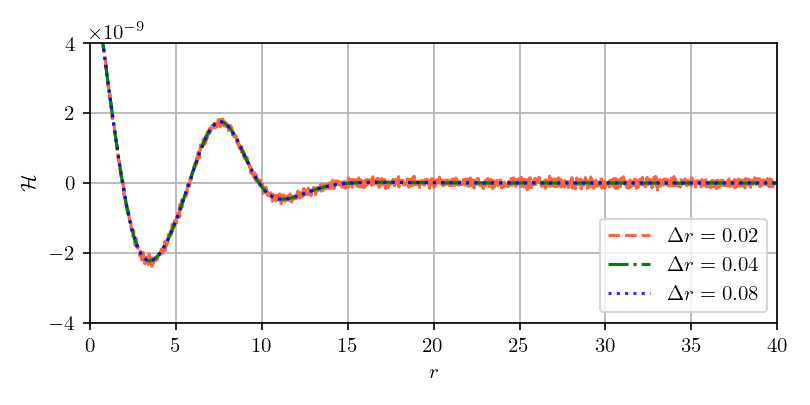}
\caption{\textit{Hamiltonian constraint for different radial resolutions $\Delta r = 0.02,0.04,0.08$. The curves corresponding to $\Delta r = 0.02$ and $\Delta r = 0.04$ have been multiplied by factors of $256$ and $16$, respectively. The excellent overlap among the curves confirms the expected fourth-order convergence of the Runge–Kutta method.}}
\label{hamiltonian-converge}
\end{figure}

We also find some differences in the frequencies $\omega$ reported in that reference.  We have not been able to understand completely the origin of these differences, but we are confident in the reliability of our results since we have performed extensive convergence tests of our code. As an example, Figure~\ref{hamiltonian-converge} shows the radial profile of the Hamiltonian constraint $\mathcal{H} = \mathper{a}' - \mathper{a} [(1-\mathper{a})/r + 8\pi r \mathper{a} \rho]$ for three different radial resolutions $\Delta r=0.02,0.04,0.08$, for the particular case with $q=0.3$ and $\psi_0=0.05$. Since $\mathcal{H} \equiv 0$, the small deviations reflect numerical truncation errors consistent with a fourth-order Runge–Kutta method. Although all our reported results were carried out using a smaller grid spacing of $\Delta r = 0.01$, this case is not shown in the plot as it already reaches the machine precision limit (for double precision numbers), and even if the Hamiltonian constraint does become even smaller it is very noisy due to the random nature of the round-off errors.


\begin{figure}[h]
    \centering
    \begin{subfigure}[b]{0.49\textwidth}
        \centering
        \includegraphics[width=\textwidth]{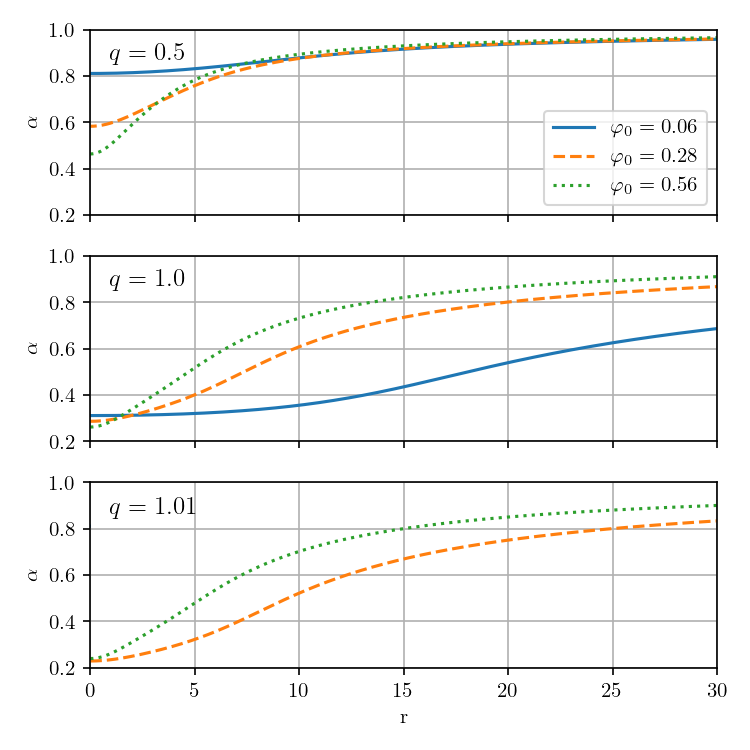}
        \label{sol-lapso}
    \end{subfigure}
    \hfill
    \begin{subfigure}[b]{0.49\textwidth}
        \centering
        \includegraphics[width=\textwidth]{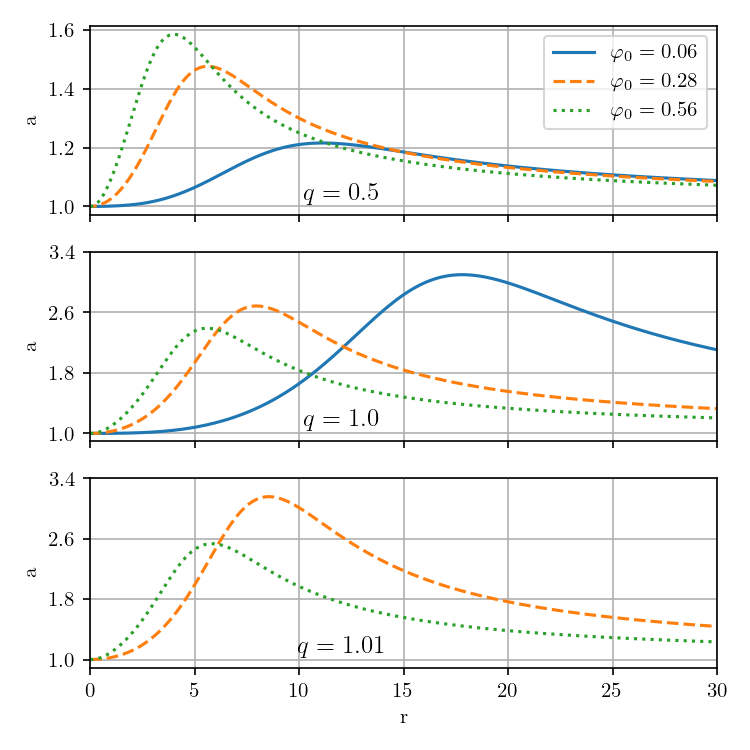}
        \label{sol-a}
    \end{subfigure}
    \vspace{-0.6cm}
    \caption{\textit{Solutions for geometric variables: lapse $\alpha(r)$ (left panel) and metric $\mathper{a(r)}$ (right panel), for three values of charge $q=0.5,1.0,1.01$ and three values of the central potential $\varphi_0=0.06, 0.28, 0.56$.}}
    \label{var-geo}
\end{figure}

\subsection{Some particular configurations}

In this section we will present some particular solutions obtained for the physical variables of our system, i.e. $\{\alpha, \mathper{a}, \varphi, x, \phi, E\}$, as well as $\mathcal{E}$, $\rho$ and $\rho_Q$, in such a way that we can study their properties. Considering the two free parameters, namely $q$ and $\varphi_0$, we have chosen to present solutions corresponding to charges $q=0.5,1.0,1.01$, that is a subcritical charge, the critical charge itself, and a supercritical one; as for the central potential, in all cases the solutions we choose correspond to $\varphi_0 = 0.06, 0.28, 0.56$. These values of $\varphi_0$ are chosen since, for the particular case of $q = 0.5$, they correspond to a bound solution with $E_B<0$ to the left of the minimum of $E_B$ (expected to be stable), a bound solution to the right of the minimum of $E_B$ (expected to be unstable), and a gravitationally unbound solution with $E_B>0$. 

\begin{figure}[h]
    \centering
    \begin{subfigure}[b]{0.70\textwidth}
        \centering
        \includegraphics[width=\textwidth]{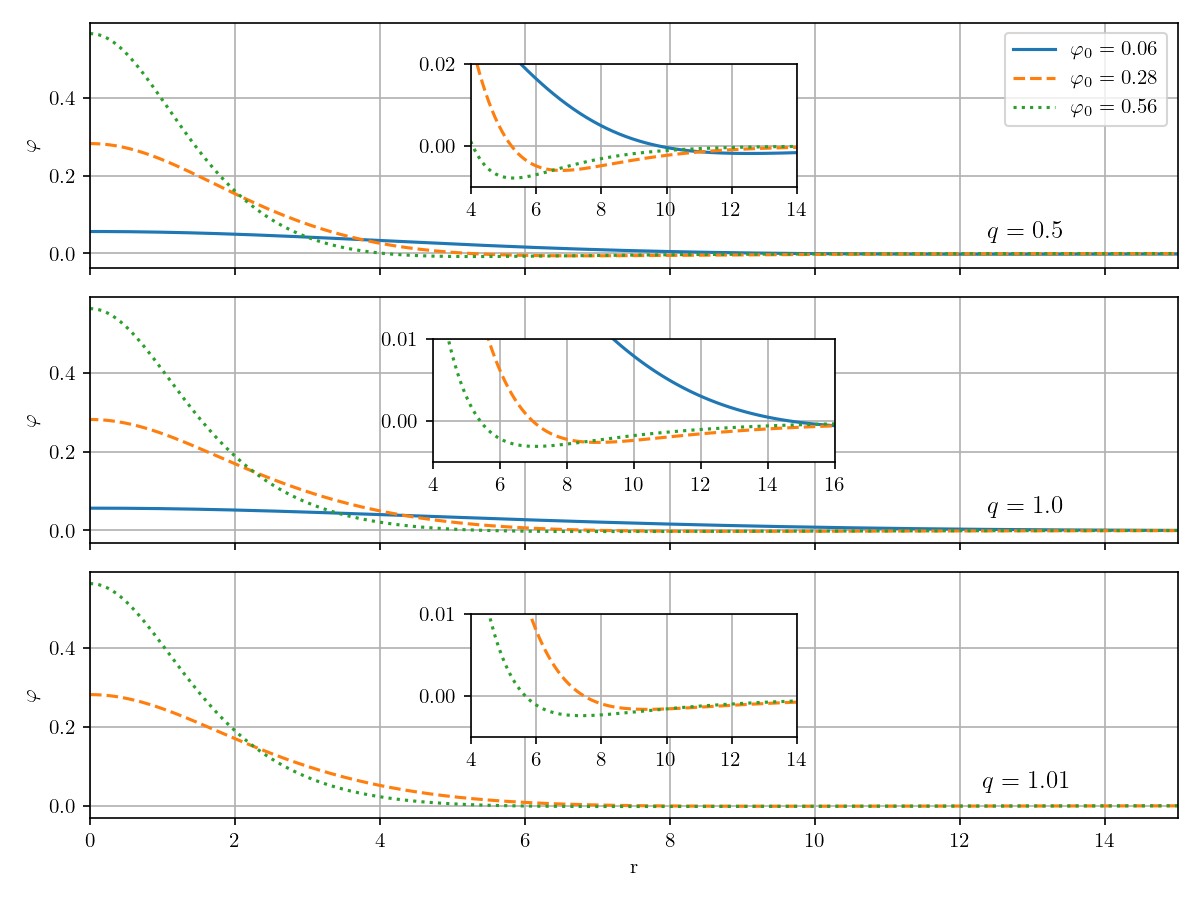}
        \label{sol-Eproca}
    \end{subfigure}
    \vspace{0.3cm} 
    \begin{subfigure}[b]{0.49\textwidth}
        \centering
        \includegraphics[width=\textwidth]{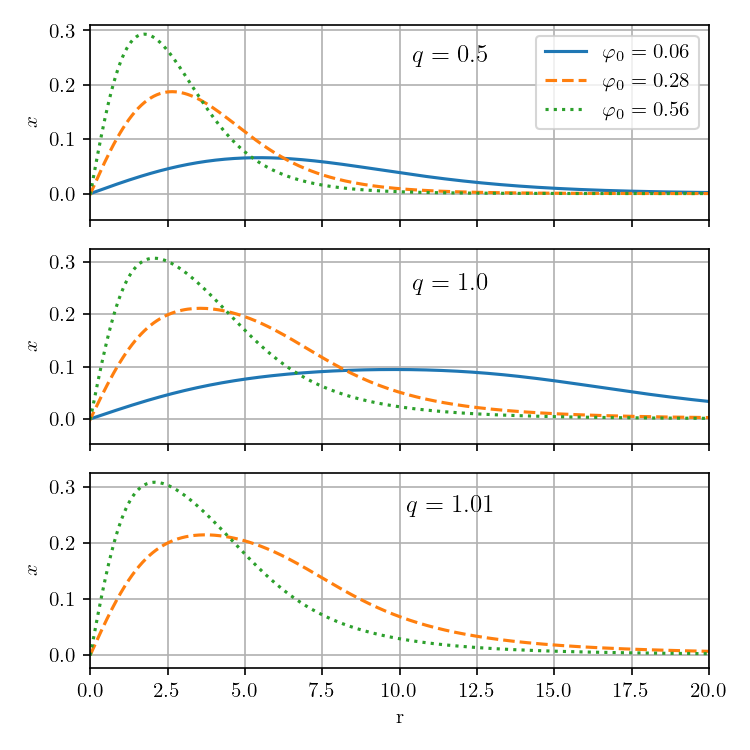}
    \end{subfigure}
    \hfill
    \begin{subfigure}[b]{0.49\textwidth}
        \centering
        \includegraphics[width=\textwidth]{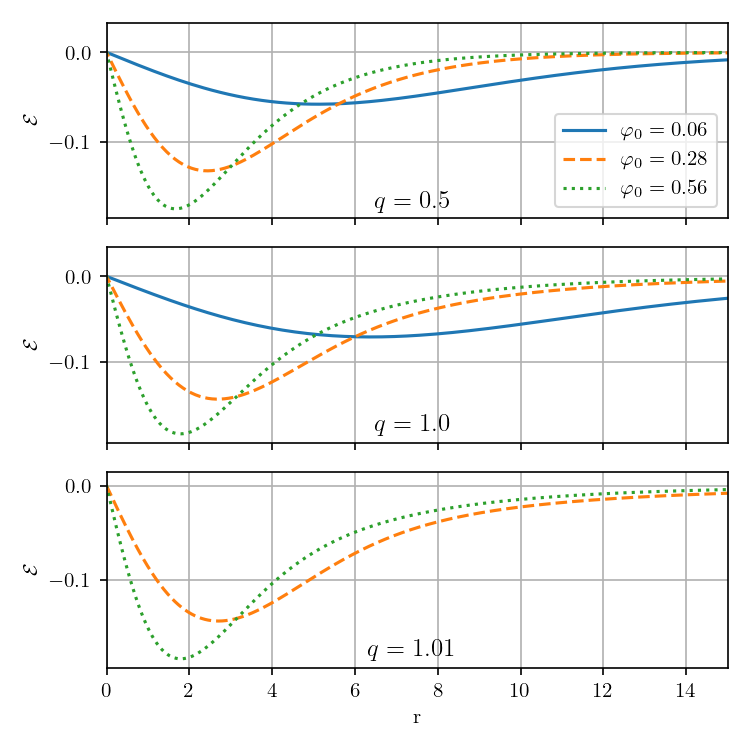}
    \end{subfigure}
    \vspace{-0.6cm}
    \caption{\textit{Solutions for the different variables of the Proca field: scalar potential $\varphi(r)$ (upper panel), vector potential $x(r)$ (lower-left panel), and electric field $\mathcal{E}$(r) (lower-right panel), for three values of the charge $q=0.5,1.0,1.01$ and three values of the central potential $\varphi_0=0.06, 0.28, 0.56$.}}
    \label{var-Proca}
\end{figure}

In the left panel of Figure~\ref{var-geo} we show the lapse function $\alpha(r)$. There is no curve for the particular case with $q=1.01$ and $\varphi_0 = 0.06$ since such a solution does not exist (it is outside the allowed range); this is why this curve is absent in the third panel of this figure and those that follow.  We verify that $\alpha \rightarrow 1$ at infinity and that $\alpha'(0)=0$. We observe that as the charge increases, the central lapse becomes smaller. On the other hand, the lapse at the origin is larger for smaller central potentials.

The right panel of Figure~\ref{var-geo} shows the curves for the solutions of the radial metric component $\mathper{a}(r)$. All of these plots show a bell shape with a well-defined maximum, whose value and position depend on a specific combination of charge and central potential. We verify that $\mathper{a}(0)=1$, $\mathper{a}'(0)=0$, and $a \rightarrow 1$ at infinity. Given a value of the charge, as we consider larger central potentials the maxima of $\mathper{a(r)}$ are found closer to the center, indicating a more compact configuration. Also, we observe that a larger value of $q$ implies larger values of $\mathper{a}(r)$.


In the upper panel of Figure \ref{var-Proca} we can see the curves representing the solutions for the Proca scalar potential $\varphi(r)$. We verify that $\varphi'(0)=0$ and that $\varphi \rightarrow 0$ at infinity. In each of the panels, we have zoomed in to observe the presence of a node in $\varphi(r)$, which corresponds to the point where the curve crosses the horizontal axis. We have found that, regardless of the value of the charge, this node in $\varphi(r)$ is always present. It is important to mention here that a recent study carried out in~\cite{Herdeiro:2024a} shows that the presence of nodes indicates that these configurations, although representing the lowest energy solutions in spherical symmetry, do not correspond to the true ground states. In that study, it was shown that the spherically symmetric Proca stars migrate to an axisymmetric prolate configuration that does not present any nodes in its scalar potential. Since our results indicate that the charge $q$ does not alter the number of nodes of our solutions, we expect that the spherically symmetric CPS will not correspond to the ground state either.

The lower left panel of Figure~\ref{var-Proca} shows the curves for the radial component of the vector potential $x(r)$. We verify that $x(0)=0$ and that $x \rightarrow 0$ at infinity. These curves again show a bell shape with a well-defined maximum. It is important to note that these curves do not have a node.
In the lower right panel of the figure, we present the solution curves of the Proca electric field $\mathcal{E}(r)$. The fact that $\mathcal{E}$ takes negative values means that the Proca electric field points toward the center of the star.  We verify that $E(0)=0$ and that $E \rightarrow 0$ at infinity. 

\begin{figure}[h]
    \centering
    \begin{subfigure}[b]{0.49\textwidth}
        \centering
        \includegraphics[width=\textwidth]{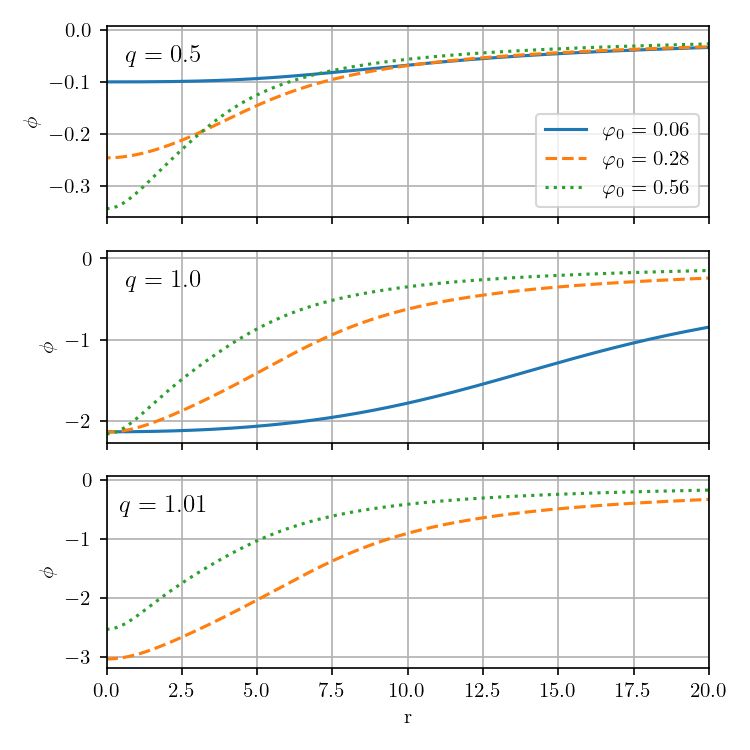}
    \end{subfigure}
    \begin{subfigure}[b]{0.49\textwidth}
        \centering
        \includegraphics[width=\textwidth]{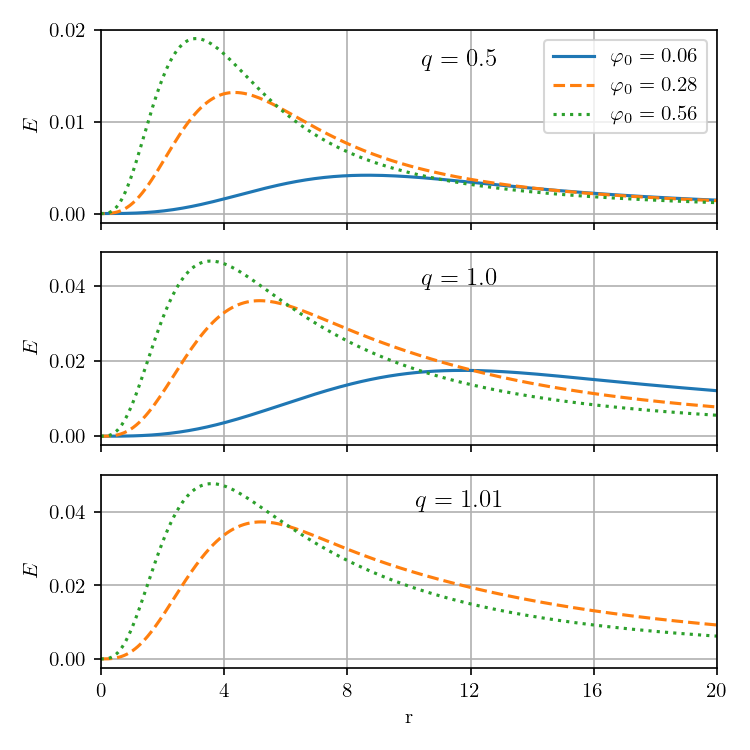}
    \end{subfigure}
    \vspace{-0.6cm}
    \caption{\textit{Solutions for the different variables of the Maxwell field: scalar potential $\phi(r)$ (left panel) and electric field $E(r)$ (right panel), for three values of the charge $q=0.5,1.0,1.01$ and three values of the central potential $\varphi_0=0.06, 0.28, 0.56$.}}
    \label{var-Maxwell}
\end{figure}

Figure~(\ref{var-Maxwell}) shows the curves corresponding to the Maxwell field. The left panel shows the scalar potential $\phi(r)$. First, we verify that $\phi'(0)=0$ and $\phi \rightarrow 0$ at infinity. The values of $\phi$ are always negative with a minimum at the origin, which ensures that the expression $\omega-q\alpha\phi$ that appears as a denominator in many of our equations never vanishes. The right panel of the figure shows the Maxwell electric field $E(r)$. We verify that $E(0)=0$, $E \rightarrow 0$ at infinity, and $E'(0)=0$. The latter is quite interesting, and is a consequence of the fact that the central charge density is also zero at the origin, as we will see below.

\begin{figure}[h]
    \centering
    \begin{subfigure}[b]{0.49\textwidth}
        \centering
        \includegraphics[width=\textwidth]{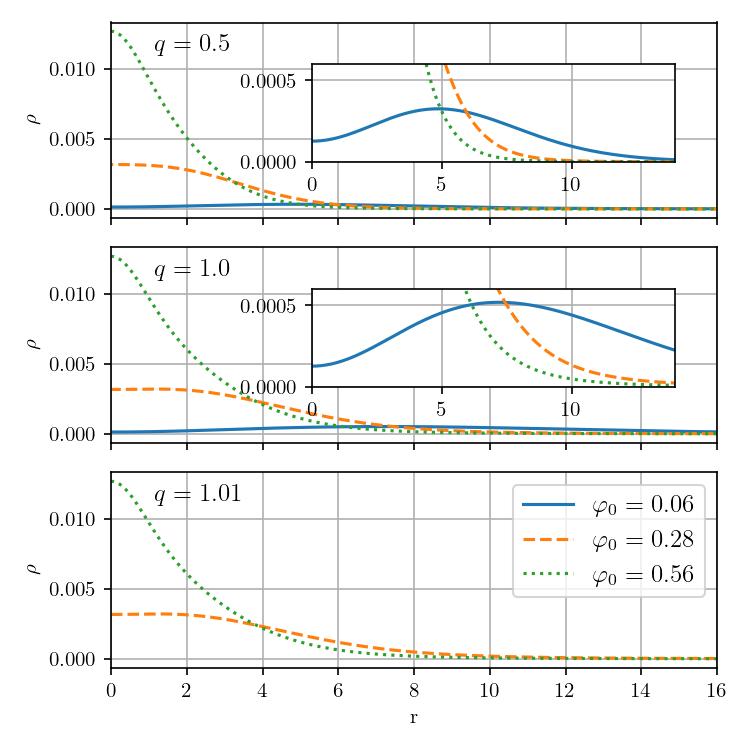}
    \end{subfigure}
    \hspace{-0.3cm}
    \begin{subfigure}[b]{0.49\textwidth}
        \centering
        \includegraphics[width=\textwidth]{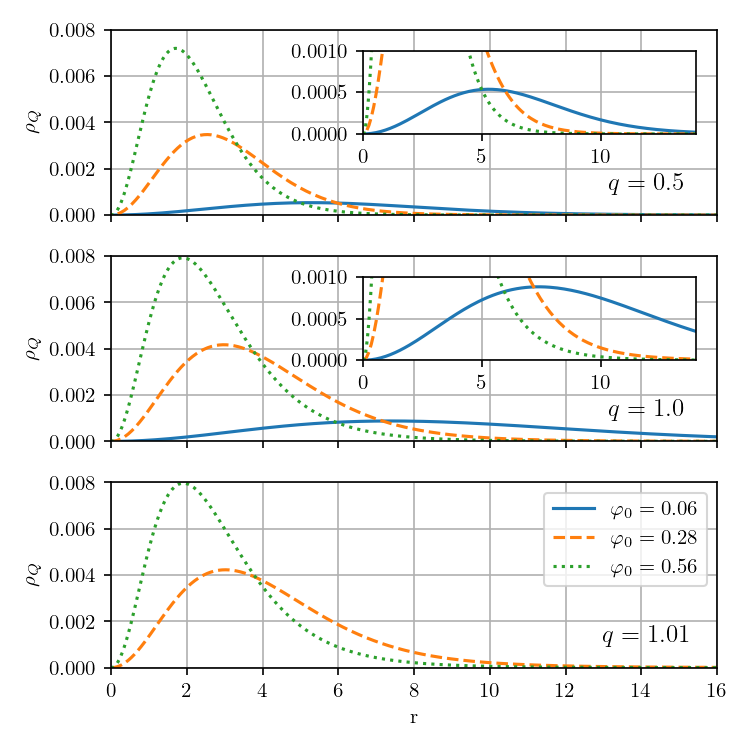}
    \end{subfigure}
    \vspace{-0.6cm}
    \caption{\textit{Energy density $\rho(r)$ (left panel) and charge density $\rho_Q(r)$ for three values of the charge $q=0.5,1.0,1.01$ and three values of the central potential $\varphi_0=0.06, 0.28, 0.56$.} }
    \label{var-rho}
\end{figure}

Finally, in Figure~\ref{var-rho}  we show plots of the energy density $\rho(r)$ (left panel) and the charge density $\rho_Q(r)$ (right panel). It is interesting that stars with large central potentials have their maximum energy density at their center, but stars with low central potentials do not. In particular, stars with $\varphi_0 = 0.06$ and $q=0.5,1.0$ have a maximum energy density away from the center, as can be seen in the insets.  Notice that the charge density at the center of the star is always zero which, as we mentioned above, explains the fact that the derivative of the Maxwell electric field $E(r)$ at that point is also zero.  Also, for stars with $\varphi_0 = 0.06$ the maximum charge density $\rho_Q$ is reached very close to the point with the maximum energy density $\rho$  as well. This appears to happen with stars with low central energy density, that is, those in the stable branch.

\section{Discussion and conclusions} 

In this paper we have considered the case of self-gravitating electrically charged Proca stars in spherical symmetry, by finding solutions to the Einstein-Maxwell-Proca (EMP) system in a static spherically symmetric spacetime. In order to construct our solutions we have first derived the general equations of the EMP system in the 3+1 formalism of general relativity. We further reduced this system of equations by considering the particular case of spherical symmetry, finding evolution and constraint equations for the Proca and Maxwell potentials and fields (\ref{ev-varphi-3+1}-\ref{cons-Em-3+1}). A second reduction of the system was obtained by particularizing it to the case of a harmonically oscillating Proca field and a static spacetime, this led us to a final system of six non-linear radial equations (\ref{eq-rad-a}-\ref{eq-rad-E}) which, given a specific value of the Proca potential at the origin $\varphi_0$, constitute an eigenvalue problem for the oscillation frequency $\omega$.

For the case with no electric charge, $q=0$, we have been able to reproduce previously known results.  We have also considered families with different values of the charge parameter $q$. Perhaps the most remarkable aspect of our work was that we not only found solutions for charges below the critical value $q_c=m$, but we have also found solutions for supercritical charges, that is, charges slightly larger than $q_c$.  Similar supercritical solutions have also been reported in~\cite{Lopez:2023phk} for the case charged boson (scalar) stars. These supercritical solutions do not cover the full range of values for $\varphi_0$, but exist only for a narrow window of values for this parameter. This allowed range of values becomes smaller as the value of $q$ increases, until no more solutions can be found for $q/m>1.04$. Although these supercritical families are characterized by having a charge parameter $q$ of the field greater than the boson mass $m$, we have found that the total integrated charge $Q$ is always less than the total mass $M$ of the star. We also find that these supercritical solutions are always gravitationally unbound, that is with a positive binding energy $E_B > 0$, and therefore can be expected to be dynamically unstable against perturbations. 

For the different charged Proca star solutions we determine the radial profiles of the lapse function $\alpha$, the radial metric $a$, the Proca potentials $\varphi$ and $x$, the Proca electric field $E$, the Maxwell potential $\phi$, the Maxwell electric field $E$, the energy density $\rho$, and charge density $\rho_Q$.  We have found that, regardless of the value of the charge $q$, the curve for the Proca scalar potential $\varphi(r)$ always presents a single node, while the curve for the Proca vector potential $x(r)$ has no nodes (solutions with higher number of nodes corresponding to excited states can also be found, but we have not considered them here). We have also found that the charge density $\rho_Q(r)$ of our solutions is always zero at the origin, which turns out to be a very special feature of these charged objects. Because of this, the radial derivative of the electric field is also zero at the center of the star.  In general, we can conclude that the role played by the charge in Proca stars is to increase their effective radius and their maximum allowed mass. Of course, this is to be expected since the charge of the Proca field generates a Coulomb repulsion that points out from the star, increasing the radius while also opposing the gravitational field which allows for higher mass solutions.

 We are currently studying the dynamical evolution of our solutions for charged Proca stars in order to understand their stability properties against perturbations, and will be reporting on these results elsewhere.
 

\acknowledgments

The authors wish to thank Axel Rangel and Carlos Joaquin for many useful discussions and comments. This work was partially supported by CONAHCYT Network Projects 376127 and 304001, and DGAPA-UNAM project IN100523.  Yahir Mio also acknowledges support from a CONAHCYT National Graduate Grant.


\bibliographystyle{apsrev4-1}
\bibliography{referencias}


\end{document}